\documentclass[%
 reprint,
 superscriptaddress,
 preprintnumbers,
 bibnotes,
 amsmath,amssymb,
 aps,
 pra,
 nofootinbib,
longbibliography
]{revtex4-2}
\usepackage{CJK}

\usepackage{mathrsfs}  

\usepackage{tcolorbox}
 \usepackage[textsize=tiny]{todonotes}

\usepackage[utf8]{inputenc}
\usepackage{graphicx}
\usetikzlibrary{shapes.misc}
\usetikzlibrary{cd}
\usepackage{dcolumn}
\usepackage{bm}
\usepackage{hyperref}
\hypersetup{
    colorlinks=true,
    linkcolor=Nblue,
    filecolor=black,      
    urlcolor=Nblue,
    citecolor=Nblue,
    breaklinks
}

\definecolor{Nblue}{RGB}{0,47,167}    
\definecolor{ku}{RGB}{144,26,30}
\definecolor{mathblue}{rgb}{0.368417, 0.506779, 0.709798}
\definecolor{mathyellow}{rgb}{0.880722, 0.611041, 0.142051}

\newcommand{\numpb}{N}
\newcommand{\numdpr}{N_R}
\newcommand{\numdpl}{N_L}

\newcommand{\fourd}{\text{\tiny (4d)}}
\newcommand{\sixd}{\text{\tiny (6d)}}
\newcommand{\eightd}{\text{\tiny (8d)}}

\newcommand{\cG}{\mathcal G}
\newcommand{\cI}{\mathcal I}

\allowdisplaybreaks

\newcommand{\generictwoloop}{%
\begin{tikzpicture}[scale=0.45,baseline=-3pt]
\draw[mathblue, very thick] (0,0) -- (-0.3,0.94);
\draw[mathblue, very thick] (0,0) -- (-0.3,-0.94);
\draw[mathblue, very thick] (0,0) -- (-0.9,0.4);
\draw[mathblue, very thick] (0,0) -- (-0.9,-0.4);
\draw[mathblue, very thick] (0,0) -- (0.55,0.83);
\draw[mathblue, very thick] (0,0) -- (0.55,-0.83);
\draw[mathblue, very thick] (0,0) -- (0.99,-0.1);
\filldraw[mathblue] (0,0) circle (15pt);
\filldraw[white] (-0.17,-0.17) circle (4.5pt);
\filldraw[white] (0.17,0.17) circle (4.5pt);
\filldraw[mathblue] (0.75,0.13) circle (1.5pt);
\filldraw[mathblue] (0.72,0.32) circle (1.5pt);
\filldraw[mathblue] (0.6,0.48) circle (1.5pt);
\end{tikzpicture}
}

\newcommand{\Db}{%
\begin{tikzpicture}[scale=0.32,baseline=-2.5pt]
\draw[mathblue, line width=1pt] (0,-0.5)--(0,0.5);
\draw[mathblue, line width=1pt] (0,-0.5)--(1,-0.5);
\draw[mathblue, line width=1pt] (0,-0.5)--(-1,-0.5);
\draw[mathblue, line width=1pt] (0,0.5)--(1,0.5);
\draw[mathblue, line width=1pt] (0,0.5)--(-1,0.5);
\draw[mathblue, line width=1pt] (1,0.5)--(1,-0.5);
\draw[mathblue, line width=1pt] (-1,0.5)--(-1,-0.5);
\draw[mathblue, line width=1pt] (0,0.5) -- (0.28,0.96);
\draw[mathblue, line width=1pt] (0,0.5) -- (-0.28,0.96);
\filldraw[mathblue] (-0.14,0.96) circle (1pt);
\filldraw[mathblue] (0,1.0) circle (1pt);
\filldraw[mathblue] (0.14,0.96) circle (1pt);
\draw[mathblue, line width=1pt] (0,-0.5) -- (0.28,-0.96);
\draw[mathblue, line width=1pt] (0,-0.5) -- (-0.28,-0.96);
\filldraw[mathblue] (-0.14,-0.96) circle (1pt);
\filldraw[mathblue] (0,-1.0) circle (1pt);
\filldraw[mathblue] (0.14,-0.96) circle (1pt);
\draw[mathblue, line width=1pt] (1,0.5) -- (1.5,0.6);
\draw[mathblue, line width=1pt] (1,0.5) -- (1.1,1.0);
\filldraw[mathblue] (1.2,0.9) circle (1pt);
\filldraw[mathblue] (1.32,0.82) circle (1pt);
\filldraw[mathblue] (1.4,0.7) circle (1pt);
\draw[mathblue, line width=1pt] (-1,0.5) -- (-1.5,0.6);
\draw[mathblue, line width=1pt] (-1,0.5) -- (-1.1,1.0);
\filldraw[mathblue] (-1.2,0.9) circle (1pt);
\filldraw[mathblue] (-1.32,0.82) circle (1pt);
\filldraw[mathblue] (-1.4,0.7) circle (1pt);
\draw[mathblue, line width=1pt] (1,-0.5) -- (1.5,-0.6);
\draw[mathblue, line width=1pt] (1,-0.5) -- (1.1,-1.0);
\filldraw[mathblue] (1.2,-0.9) circle (1pt);
\filldraw[mathblue] (1.32,-0.82) circle (1pt);
\filldraw[mathblue] (1.4,-0.7) circle (1pt);
\draw[mathblue, line width=1pt] (-1,-0.5) -- (-1.5,-0.6);
\draw[mathblue, line width=1pt] (-1,-0.5) -- (-1.1,-1.0);
\filldraw[mathblue] (-1.2,-0.9) circle (1pt);
\filldraw[mathblue] (-1.32,-0.82) circle (1pt);
\filldraw[mathblue] (-1.4,-0.7) circle (1pt);
\end{tikzpicture}
}

\newcommand{\Pb}{
\begin{tikzpicture}[scale=0.32,baseline=-2.5pt]
\draw[mathblue, line width=1pt] (0,-0.5)--(0,0.5);
\draw[mathblue, line width=1pt] (0,-0.5)--(-0.95,-0.8);
\draw[mathblue, line width=1pt] (0,0.5)--(-0.95,0.8);
\draw[mathblue, line width=1pt] (0,-0.5)--(1,-0.5);
\draw[mathblue, line width=1pt] (0,0.5)--(1,0.5);
\draw[mathblue, line width=1pt] (-1.55,0)--(-0.95,-0.8);
\draw[mathblue, line width=1pt] (-1.55,0)--(-0.95,0.8);
\draw[mathblue, line width=1pt] (1,0.5)--(1,-0.5);
\draw[mathblue, line width=1pt] (0,0.5) -- (0.35,0.91);
\draw[mathblue, line width=1pt] (0,0.5) -- (-0.21,1.00);
\filldraw[mathblue] (-0.07,0.98) circle (1pt);
\filldraw[mathblue] (0.08,0.99) circle (1pt);
\filldraw[mathblue] (0.21,0.93) circle (1pt);
\draw[mathblue, line width=1pt] (0,-0.5) -- (0.35,-0.91);
\draw[mathblue, line width=1pt] (0,-0.5) -- (-0.21,-1.00);
\filldraw[mathblue] (-0.07,-0.98) circle (1pt);
\filldraw[mathblue] (0.08,-0.99) circle (1pt);
\filldraw[mathblue] (0.21,-0.93) circle (1pt);
\draw[mathblue, line width=1pt] (-1.55,0) -- (-2.01,0.28);
\draw[mathblue, line width=1pt] (-1.55,0) -- (-2.01,-0.28);
\filldraw[mathblue] (-2.01,-0.14) circle (1pt);
\filldraw[mathblue] (-2.05,0) circle (1pt);
\filldraw[mathblue] (-2.01,0.14) circle (1pt);
\draw[mathblue, line width=1pt] (1,0.5) -- (1.5,0.6);
\draw[mathblue, line width=1pt] (1,0.5) -- (1.1,1.0);
\filldraw[mathblue] (1.2,0.9) circle (1pt);
\filldraw[mathblue] (1.32,0.82) circle (1pt);
\filldraw[mathblue] (1.4,0.7) circle (1pt);
\draw[mathblue, line width=1pt] (1,-0.5) -- (1.5,-0.6);
\draw[mathblue, line width=1pt] (1,-0.5) -- (1.1,-1.0);
\filldraw[mathblue] (1.2,-0.9) circle (1pt);
\filldraw[mathblue] (1.32,-0.82) circle (1pt);
\filldraw[mathblue] (1.4,-0.7) circle (1pt);
\draw[mathblue, line width=1pt] (-0.95,-0.8) -- (-0.83,-1.32);
\draw[mathblue, line width=1pt] (-0.95,-0.8) -- (-1.36,-1.15);
\filldraw[mathblue] (-0.96,-1.28) circle (1pt);
\filldraw[mathblue] (-1.10,-1.27) circle (1pt);
\filldraw[mathblue] (-1.23,-1.19) circle (1pt);
\draw[mathblue, line width=1pt] (-0.95,0.8) -- (-0.83,1.32);
\draw[mathblue, line width=1pt] (-0.95,0.8) -- (-1.36,1.15);
\filldraw[mathblue] (-0.96,1.28) circle (1pt);
\filldraw[mathblue] (-1.10,1.27) circle (1pt);
\filldraw[mathblue] (-1.23,1.19) circle (1pt);
\end{tikzpicture}
}

\newcommand{\Ddp}{
\begin{tikzpicture}[scale=0.32,baseline=-2.5pt]
\draw[mathblue, line width=1pt] (0,-0.5)--(0,0.5);
\draw[mathblue, line width=1pt] (0,-0.5)--(-0.95,-0.8);
\draw[mathblue, line width=1pt] (0,-0.5)--(0.95,-0.8);
\draw[mathblue, line width=1pt] (0,0.5)--(-0.95,0.8);
\draw[mathblue, line width=1pt] (0,0.5)--(0.95,0.8);
\draw[mathblue, line width=1pt] (-1.55,0)--(-0.95,-0.8);
\draw[mathblue, line width=1pt] (1.55,0)--(0.95,-0.8);
\draw[mathblue, line width=1pt] (-1.55,0)--(-0.95,0.8);
\draw[mathblue, line width=1pt] (1.55,0)--(0.95,0.8);
\draw[mathblue, line width=1pt] (0,0.5) -- (0.28,0.96);
\draw[mathblue, line width=1pt] (0,0.5) -- (-0.28,0.96);
\filldraw[mathblue] (-0.14,0.96) circle (1pt);
\filldraw[mathblue] (0,1.0) circle (1pt);
\filldraw[mathblue] (0.14,0.96) circle (1pt);
\draw[mathblue, line width=1pt] (0,-0.5) -- (0.28,-0.96);
\draw[mathblue, line width=1pt] (0,-0.5) -- (-0.28,-0.96);
\filldraw[mathblue] (-0.14,-0.96) circle (1pt);
\filldraw[mathblue] (0,-1.0) circle (1pt);
\filldraw[mathblue] (0.14,-0.96) circle (1pt);
\draw[mathblue, line width=1pt] (-1.55,0) -- (-2.01,0.28);
\draw[mathblue, line width=1pt] (-1.55,0) -- (-2.01,-0.28);
\filldraw[mathblue] (-2.01,-0.14) circle (1pt);
\filldraw[mathblue] (-2.05,0) circle (1pt);
\filldraw[mathblue] (-2.01,0.14) circle (1pt);
\draw[mathblue, line width=1pt] (1.55,0) -- (2.01,0.28);
\draw[mathblue, line width=1pt] (1.55,0) -- (2.01,-0.28);
\filldraw[mathblue] (2.01,-0.14) circle (1pt);
\filldraw[mathblue] (2.05,0) circle (1pt);
\filldraw[mathblue] (2.01,0.14) circle (1pt);
\draw[mathblue, line width=1pt] (-0.95,-0.8) -- (-0.83,-1.32);
\draw[mathblue, line width=1pt] (-0.95,-0.8) -- (-1.36,-1.15);
\filldraw[mathblue] (-0.96,-1.28) circle (1pt);
\filldraw[mathblue] (-1.10,-1.27) circle (1pt);
\filldraw[mathblue] (-1.23,-1.19) circle (1pt);
\draw[mathblue, line width=1pt] (-0.95,0.8) -- (-0.83,1.32);
\draw[mathblue, line width=1pt] (-0.95,0.8) -- (-1.36,1.15);
\filldraw[mathblue] (-0.96,1.28) circle (1pt);
\filldraw[mathblue] (-1.10,1.27) circle (1pt);
\filldraw[mathblue] (-1.23,1.19) circle (1pt);
\draw[mathblue, line width=1pt] (0.95,-0.8) -- (0.83,-1.32);
\draw[mathblue, line width=1pt] (0.95,-0.8) -- (1.36,-1.15);
\filldraw[mathblue] (0.96,-1.28) circle (1pt);
\filldraw[mathblue] (1.10,-1.27) circle (1pt);
\filldraw[mathblue] (1.23,-1.19) circle (1pt);
\draw[mathblue, line width=1pt] (0.95,0.8) -- (0.83,1.32);
\draw[mathblue, line width=1pt] (0.95,0.8) -- (1.36,1.15);
\filldraw[mathblue] (0.96,1.28) circle (1pt);
\filldraw[mathblue] (1.10,1.27) circle (1pt);
\filldraw[mathblue] (1.23,1.19) circle (1pt);
\end{tikzpicture}
}

\newcommand{\SB}{%
\begin{tikzpicture}[scale=0.2,baseline=-2.5pt]
\draw (0,-0.5)--(0,0.5);
\draw (0,-0.5)--(-1,-0.5);
\draw (0,0.5)--(-1,0.5);
\draw (-1,0.5)--(-1,-0.5);
\draw (0,0.5) -- (0.42,0.78);
\draw (0,0.5) -- (0.28,0.92);
\draw (-1,0.5) -- (-1.42,0.78);
\draw (-1,0.5) -- (-1.28,0.92);
\draw (0,-0.5) -- (0.42,-0.78);
\draw (0,-0.5) -- (0.28,-0.92);
\draw (-1,-0.5) -- (-1.42,-0.78);
\draw (-1,-0.5) -- (-1.28,-0.92);
\end{tikzpicture}
}

\newcommand{\DB}{%
\begin{tikzpicture}[scale=0.17,baseline=-2.5pt]
\draw (0,-0.5)--(0,0.5);
\draw (0,-0.5)--(1,-0.5);
\draw (0,-0.5)--(-1,-0.5);
\draw (0,0.5)--(1,0.5);
\draw (0,0.5)--(-1,0.5);
\draw (1,0.5)--(1,-0.5);
\draw (-1,0.5)--(-1,-0.5);
\draw (0,0.5) -- (-0.1,1.0);
\draw (0,0.5) -- (0.1,1.0);
\draw (0,-0.5) -- (-0.1,-1.0);
\draw (0,-0.5) -- (0.1,-1.0);
\draw (1,0.5) -- (1.42,0.78);
\draw (1,0.5) -- (1.28,0.92);
\draw (-1,0.5) -- (-1.42,0.78);
\draw (-1,0.5) -- (-1.28,0.92);
\draw (1,-0.5) -- (1.42,-0.78);
\draw (1,-0.5) -- (1.28,-0.92);
\draw (-1,-0.5) -- (-1.42,-0.78);
\draw (-1,-0.5) -- (-1.28,-0.92);
\end{tikzpicture}
}

\newcommand{\PB}{
\begin{tikzpicture}[scale=0.17,baseline=-2.5pt]
\draw (0,-0.5)--(0,0.5);
\draw (0,-0.5)--(-0.95,-0.8);
\draw (0,0.5)--(-0.95,0.8);
\draw (0,-0.5)--(1,-0.5);
\draw (0,0.5)--(1,0.5);
\draw (-1.55,0)--(-0.95,-0.8);
\draw (-1.55,0)--(-0.95,0.8);
\draw (1,0.5)--(1,-0.5);
\draw (0,0.5) -- (-0.1,1.0);
\draw (0,0.5) -- (0.1,1.0);
\draw (0,-0.5) -- (-0.1,-1.0);
\draw (0,-0.5) -- (0.1,-1.0);
\draw (-1.55,0) -- (-2.05,0.1);
\draw (-1.55,0) -- (-2.05,-0.1);
\draw (1,0.5) -- (1.42,0.78);
\draw (1,0.5) -- (1.28,0.92);
\draw (1,-0.5) -- (1.42,-0.78);
\draw (1,-0.5) -- (1.28,-0.92);
\draw (-0.95,-0.8) -- (-1.20,-1.25);
\draw (-0.95,-0.8) -- (-1.00,-1.31);
\draw (-0.95,0.8) -- (-1.20,1.25);
\draw (-0.95,0.8) -- (-1.00,1.31);
\node[mathblue!50!black] at (-0.7,0) {\tiny $\numpb$};
\end{tikzpicture}
}

\newcommand{\PBnoN}{
\begin{tikzpicture}[scale=0.17,baseline=-2.5pt]
\draw (0,-0.5)--(0,0.5);
\draw (0,-0.5)--(-0.95,-0.8);
\draw (0,0.5)--(-0.95,0.8);
\draw (0,-0.5)--(1,-0.5);
\draw (0,0.5)--(1,0.5);
\draw (-1.55,0)--(-0.95,-0.8);
\draw (-1.55,0)--(-0.95,0.8);
\draw (1,0.5)--(1,-0.5);
\draw (0,0.5) -- (-0.1,1.0);
\draw (0,0.5) -- (0.1,1.0);
\draw (0,-0.5) -- (-0.1,-1.0);
\draw (0,-0.5) -- (0.1,-1.0);
\draw (-1.55,0) -- (-2.05,0.1);
\draw (-1.55,0) -- (-2.05,-0.1);
\draw (1,0.5) -- (1.42,0.78);
\draw (1,0.5) -- (1.28,0.92);
\draw (1,-0.5) -- (1.42,-0.78);
\draw (1,-0.5) -- (1.28,-0.92);
\draw (-0.95,-0.8) -- (-1.20,-1.25);
\draw (-0.95,-0.8) -- (-1.00,-1.31);
\draw (-0.95,0.8) -- (-1.20,1.25);
\draw (-0.95,0.8) -- (-1.00,1.31);
\end{tikzpicture}
}

\newcommand{\PBone}{
\begin{tikzpicture}[scale=0.17,baseline=-2.5pt]
\draw (0,-0.5)--(0,0.5);
\draw (0,-0.5)--(-0.95,-0.8);
\draw (0,0.5)--(-0.95,0.8);
\draw (0,-0.5)--(1,-0.5);
\draw (0,0.5)--(1,0.5);
\draw (-1.55,0)--(-0.95,-0.8);
\draw (-1.55,0)--(-0.95,0.8);
\draw (1,0.5)--(1,-0.5);
\draw (0,0.5) -- (-0.1,1.0);
\draw (0,0.5) -- (0.1,1.0);
\draw (0,-0.5) -- (-0.1,-1.0);
\draw (0,-0.5) -- (0.1,-1.0);
\draw (-1.55,0) -- (-2.05,0.1);
\draw (-1.55,0) -- (-2.05,-0.1);
\draw (1,0.5) -- (1.42,0.78);
\draw (1,0.5) -- (1.28,0.92);
\draw (1,-0.5) -- (1.42,-0.78);
\draw (1,-0.5) -- (1.28,-0.92);
\draw (-0.95,-0.8) -- (-1.20,-1.25);
\draw (-0.95,-0.8) -- (-1.00,-1.31);
\draw (-0.95,0.8) -- (-1.20,1.25);
\draw (-0.95,0.8) -- (-1.00,1.31);
\filldraw (0,0) circle (4pt);
\filldraw (0.5,0.5) circle (4pt);
\end{tikzpicture}
}

\newcommand{\PBtwo}{
\begin{tikzpicture}[scale=0.17,baseline=-2.5pt]
\draw (0,-0.5)--(0,0.5);
\draw (0,-0.5)--(-0.95,-0.8);
\draw (0,0.5)--(-0.95,0.8);
\draw (0,-0.5)--(1,-0.5);
\draw (0,0.5)--(1,0.5);
\draw (-1.55,0)--(-0.95,-0.8);
\draw (-1.55,0)--(-0.95,0.8);
\draw (1,0.5)--(1,-0.5);
\draw (0,0.5) -- (-0.1,1.0);
\draw (0,0.5) -- (0.1,1.0);
\draw (0,-0.5) -- (-0.1,-1.0);
\draw (0,-0.5) -- (0.1,-1.0);
\draw (-1.55,0) -- (-2.05,0.1);
\draw (-1.55,0) -- (-2.05,-0.1);
\draw (1,0.5) -- (1.42,0.78);
\draw (1,0.5) -- (1.28,0.92);
\draw (1,-0.5) -- (1.42,-0.78);
\draw (1,-0.5) -- (1.28,-0.92);
\draw (-0.95,-0.8) -- (-1.20,-1.25);
\draw (-0.95,-0.8) -- (-1.00,-1.31);
\draw (-0.95,0.8) -- (-1.20,1.25);
\draw (-0.95,0.8) -- (-1.00,1.31);
\filldraw (0,0) circle (4pt);
\filldraw (1,0) circle (4pt);
\end{tikzpicture}
}

\newcommand{\PBthree}{
\begin{tikzpicture}[scale=0.17,baseline=-2.5pt]
\draw (0,-0.5)--(0,0.5);
\draw (0,-0.5)--(-0.95,-0.8);
\draw (0,0.5)--(-0.95,0.8);
\draw (0,-0.5)--(1,-0.5);
\draw (0,0.5)--(1,0.5);
\draw (-1.55,0)--(-0.95,-0.8);
\draw (-1.55,0)--(-0.95,0.8);
\draw (1,0.5)--(1,-0.5);
\draw (0,0.5) -- (-0.1,1.0);
\draw (0,0.5) -- (0.1,1.0);
\draw (0,-0.5) -- (-0.1,-1.0);
\draw (0,-0.5) -- (0.1,-1.0);
\draw (-1.55,0) -- (-2.05,0.1);
\draw (-1.55,0) -- (-2.05,-0.1);
\draw (1,0.5) -- (1.42,0.78);
\draw (1,0.5) -- (1.28,0.92);
\draw (1,-0.5) -- (1.42,-0.78);
\draw (1,-0.5) -- (1.28,-0.92);
\draw (-0.95,-0.8) -- (-1.20,-1.25);
\draw (-0.95,-0.8) -- (-1.00,-1.31);
\draw (-0.95,0.8) -- (-1.20,1.25);
\draw (-0.95,0.8) -- (-1.00,1.31);
\filldraw (0,0) circle (4pt);
\filldraw (0.5,-0.5) circle (4pt);
\end{tikzpicture}
}

\newcommand{\DP}{
\begin{tikzpicture}[scale=0.17,baseline=-2.5pt]
\draw (0,-0.5)--(0,0.5);
\draw (0,-0.5)--(-0.95,-0.8);
\draw (0,-0.5)--(0.95,-0.8);
\draw (0,0.5)--(-0.95,0.8);
\draw (0,0.5)--(0.95,0.8);
\draw (-1.55,0)--(-0.95,-0.8);
\draw (1.55,0)--(0.95,-0.8);
\draw (-1.55,0)--(-0.95,0.8);
\draw (1.55,0)--(0.95,0.8);
\draw (0,0.5) -- (-0.1,1.0);
\draw (0,0.5) -- (0.1,1.0);
\draw (0,-0.5) -- (-0.1,-1.0);
\draw (0,-0.5) -- (0.1,-1.0);
\draw (-1.55,0) -- (-2.05,0.1);
\draw (-1.55,0) -- (-2.05,-0.1);
\draw (1.55,0) -- (2.05,0.1);
\draw (1.55,0) -- (2.05,-0.1);
\draw (-0.95,-0.8) -- (-1.20,-1.25);
\draw (-0.95,-0.8) -- (-1.00,-1.31);
\draw (0.95,-0.8) -- (1.20,-1.25);
\draw (0.95,-0.8) -- (1.00,-1.31);
\draw (-0.95,0.8) -- (-1.20,1.25);
\draw (-0.95,0.8) -- (-1.00,1.31);
\draw (0.95,0.8) -- (1.20,1.25);
\draw (0.95,0.8) -- (1.00,1.31);
\node[mathblue!50!black] at (-0.7,0) {\tiny $\numpb$};
\node[mathblue!50!black] at (0.7,0) {\tiny $\numpb$};
\end{tikzpicture}
}

\newcommand{\HEX}{
\begin{tikzpicture}[scale=0.17,baseline=-2.5pt]
\draw (-0.61,0.35)--(-0.61,-0.35);
\draw (-0.61,-0.35) -- (0,-0.7);
\draw (-0.61,0.35) -- (0,0.7);
\draw (0.61,0.35)--(0.61,-0.35);
\draw (0.61,0.35) -- (0,0.7);
\draw (0.61,-0.35) -- (0,-0.7);
\draw (0,0.7) -- (-0.1,1.2);
\draw (0,0.7) -- (0.1,1.2);
\draw (0,-0.7) -- (-0.1,-1.2);
\draw (0,-0.7) -- (0.1,-1.2);
\draw (0.61,0.35) -- (1.09,0.51);
\draw (0.61,0.35) -- (0.99,0.68);
\draw (-0.61,0.35) -- (-1.09,0.51);
\draw (-0.61,0.35) -- (-0.99,0.68);
\draw (-0.61,-0.35) -- (-1.09,-0.51);
\draw (-0.61,-0.35) -- (-0.99,-0.68);
\draw (0.61,-0.35) -- (1.09,-0.51);
\draw (0.61,-0.35) -- (0.99,-0.68);
\end{tikzpicture}
}

\newcommand{\HEP}{
\begin{tikzpicture}[scale=0.17,baseline=-2.5pt]
\draw (0.77,-0.35)--(0.77,0.35);
\draw (0.77,-0.35)--(0.22,-0.79);
\draw (0.77,0.35)--(0.22,0.79);
\draw (0.22,0.79) -- (-0.46,0.63);
\draw (0.22,-0.79) -- (-0.46,-0.63);
\draw (-0.46,-0.63) -- (-0.77,0);
\draw (-0.46,0.63) -- (-0.77,0);
\fill (0.5,-0.57) circle (0.2);
\draw (-0.77,0) -- (-1.28,0.1);
\draw (-0.77,0) -- (-1.28,-0.1);
\draw (-0.46,-0.63) -- (-0.85,-0.96);
\draw (-0.46,-0.63) -- (-0.69,-1.08);
\draw (-0.46,0.63) -- (-0.85,0.96);
\draw (-0.46,0.63) -- (-0.69,1.08);
\draw (0.22,-0.79) -- (0.23,-1.30);
\draw (0.22,-0.79) -- (0.43,-1.26);
\draw (0.22,0.79) -- (0.23,1.30);
\draw (0.22,0.79) -- (0.43,1.26);
\draw (0.77,-0.35) -- (1.18,-0.66);
\draw (0.77,-0.35) -- (1.26,-0.48);
\draw (0.77,0.35) -- (1.18,0.66);
\draw (0.77,0.35) -- (1.26,0.48);
\end{tikzpicture}
}
		   
\newcommand{\DPsix}{
\begin{tikzpicture}[scale=0.17,baseline=-2.5pt]
\draw (0,-0.5)--(0,0.5);
\draw (0,-0.5)--(-0.95,-0.8);
\draw (0,-0.5)--(0.95,-0.8);
\draw (0,0.5)--(-0.95,0.8);
\draw (0,0.5)--(0.95,0.8);
\draw (-1.55,0)--(-0.95,-0.8);
\draw (1.55,0)--(0.95,-0.8);
\draw (-1.55,0)--(-0.95,0.8);
\draw (1.55,0)--(0.95,0.8);
\draw (0,0.5) -- (-0.1,1.0);
\draw (0,0.5) -- (0.1,1.0);
\draw (0,-0.5) -- (-0.1,-1.0);
\draw (0,-0.5) -- (0.1,-1.0);
\draw (-1.55,0) -- (-2.05,0.1);
\draw (-1.55,0) -- (-2.05,-0.1);
\draw (1.55,0) -- (2.05,0.1);
\draw (1.55,0) -- (2.05,-0.1);
\draw (-0.95,-0.8) -- (-1.20,-1.25);
\draw (-0.95,-0.8) -- (-1.00,-1.31);
\draw (0.95,-0.8) -- (1.20,-1.25);
\draw (0.95,-0.8) -- (1.00,-1.31);
\draw (-0.95,0.8) -- (-1.20,1.25);
\draw (-0.95,0.8) -- (-1.00,1.31);
\draw (0.95,0.8) -- (1.20,1.25);
\draw (0.95,0.8) -- (1.00,1.31);
\fill (0,0) circle (0.2);
\end{tikzpicture}
}

\newcommand{\OCT}{
\begin{tikzpicture}[scale=0.17,baseline=-2.5pt]
\draw (-0.91,0)--(-0.64,-0.64);
\draw (0.64,0.64)--(0.91,0);
\draw (-0.64,-0.64)--(0,-0.91);
\draw (0.64,-0.64)--(0,-0.91);
\draw (-0.64,0.64)--(0,0.91);
\draw (0.64,0.64)--(0,0.91);
\draw (-0.91,0)--(-0.64,0.64);
\draw (0.91,0)--(0.64,-0.64);
\draw (0,0.91) -- (0.1,1.41);
\draw (0,0.91) -- (-0.1,1.41);
\draw (0,-0.91) -- (0.1,-1.41);
\draw (0,-0.91) -- (-0.1,-1.41);
\draw (0.91,0) -- (1.41,0.1);
\draw (0.91,0) -- (1.41,-0.1);
\draw (-0.91,0) -- (-1.41,0.1);
\draw (-0.91,0) -- (-1.41,-0.1);
\draw (0.64,0.64) -- (1.06,0.92);
\draw (0.64,0.64) -- (0.92,1.06);
\draw (-0.64,0.64) -- (-1.06,0.92);
\draw (-0.64,0.64) -- (-0.92,1.06);
\draw (0.64,-0.64) -- (1.06,-0.92);
\draw (0.64,-0.64) -- (0.92,-1.06);
\draw (-0.64,-0.64) -- (-1.06,-0.92);
\draw (-0.64,-0.64) -- (-0.92,-1.06);
\end{tikzpicture}
}

\begin{document}

\hfill \small{HU-EP-24/18-RTG $\vert$ BONN-TH-2024-09}

\begin{CJK*}{UTF8}{gbsn}

\title{All planar two-loop amplitudes in maximally supersymmetric Yang-Mills theory}

\author{Anne Spiering}
\email{spiering@physik.hu-berlin.de}
 \affiliation{Institut f\"ur Physik, Humboldt Universit\"at zu Berlin, 10117 Berlin, Germany}
 \affiliation{Niels Bohr International Academy, Niels Bohr Institute, Copenhagen University, 2100 Copenhagen \O{}, Denmark}
 \author{Matthias Wilhelm}
 \email{matthias.wilhelm@nbi.ku.dk}
 \affiliation{Niels Bohr International Academy, Niels Bohr Institute, Copenhagen University, 2100 Copenhagen \O{}, Denmark}
 \author{Chi Zhang (张驰)}
 \email{czhang@uni-bonn.de}
 \affiliation{Niels Bohr International Academy, Niels Bohr Institute, Copenhagen University, 2100 Copenhagen \O{}, Denmark}
 \affiliation{Bethe Center for Theoretical Physics, Universit\"at Bonn, 53115 Bonn, Germany}

\begin{abstract}
We calculate the general planar dual-conformally invariant double-pentagon and pentabox integrals in four dimensions.
Concretely, we derive one-fold integral representations for these elliptic integrals over polylogarithms of weight three.
These integral representations 
allow us to determine the respective symbols using consistency conditions alone.
Together with the previously calculated double-box integral, these integrals suffice to express all two-loop planar scattering amplitudes in maximally supersymmetric Yang-Mills theory in four dimensions -- which we thus calculate for any number of particles and any helicity configuration!
\end{abstract}

\maketitle

\end{CJK*}

\section{Introduction}

Since the solution of the one-loop scattering problem 
in planar maximally supersymmetric Yang-Mills ($\mathcal{N}=4$ SYM) theory~\cite{Bern:1994zx} over 30 years ago, significant advances have been made both in our computational capabilities, 
and in revealing hidden structures in general Quantum Field Theory amplitudes. 
These achievements are partly due to the full mastery over the structure of one-loop amplitudes, exemplified by techniques such as unitarity methods and the on-shell formalism~\cite{Bern:1994cg,Britto:2004ap,Britto:2005fq,ArkaniHamed:2012nw}; see e.g.\ refs.~\cite{Henn:2014yza,Elvang:2015rqa}  for a review.

While impressive progress in computing 
specific multi-loop amplitudes and Feynman integrals has been achieved, see e.g.\ ref.\ \cite{Travaglini:2022uwo}, there remains a lack of knowledge about the general two-loop scattering problem, even in planar $\mathcal N=4$ SYM theory.
The main difficulty for such computations is the appearance of functions beyond multiple polylogarithms (MPLs)~\cite{Bourjaily:2022bwx}. 
The most prominent and well-studied example is the sunrise integral~\cite{Laporta:2004rb,Adams:2013nia,Bloch:2013tra,Broedel:2017siw,Bogner:2019lfa}, which contains an elliptic curve and 
integrates to elliptic multiple polylogarithms (eMPLs)~\cite{brown2011multiple,Broedel:2017kkb}. 
Planar $\mathcal{N}=4$ SYM theory offers an excellent framework for studying scattering amplitudes due to the absence of ultraviolet divergences and full control of the infrared (IR) behaviour~\cite{Anastasiou:2003kj,Bern:2005iz}.
Notably, the first elliptic Feynman integral in this theory was identified in a specific component of the two-loop amplitude: the ten-point double-box integral~\cite{CaronHuot:2012ab,Paulos:2012nu,Nandan:2013ip,Bourjaily:2017bsb}. 
Its result in terms of eMPLs was first given in ref.\ \cite{Kristensson:2021ani} and its singularity structure was studied therein by means of symbol techniques~\cite{Goncharov:2010jf,Broedel:2018iwv,Wilhelm:2022wow}.  

In this letter, as a first step towards a comprehensive understanding of general two-loop amplitudes, 
we present a one-fold integral representation and the symbol for \emph{all two-loop amplitudes} in planar $\mathcal{N}=4$ SYM theory, regardless of the multiplicity and helicity configuration of the external particles. 
This progress is based on the known reduction of two-loop amplitudes into double-box, pentabox and double-pentagon Feynman integrals in four dimensions~\cite{Bourjaily:2015jna,Bourjaily:2017wjl}:~
\begin{align}
\label{eq: generic two loop}
 \left.\generictwoloop\right|_{\tiny \substack{\text{planar}\hspace{0.45cm}\\ \mathcal{N}\text{=4\,SYM}}}\hspace{-20pt}=\textstyle\sum_i\! a_i\Db+\!\sum_i\! b_i\,\Pb +\!\sum_i\!c_i\,\Ddp\,,
\end{align}
where the sums run over all planar momentum configurations and the coefficients can be obtained via unitarity methods.
The (twelve-point) double-box integral is the best understood of these integrals, in particular due to a first-order differential equation relating it to the one-loop hexagon integral in six dimensions \cite{Paulos:2012nu,Nandan:2013ip}; cf.\ Sec.\,\ref{sec:dbDE}. The latter was explicitly evaluated as a weight-three MPL \cite{Ren:2023tuj}, thus allowing for a one-fold integral representation over polylogarithms for the double box. 
This one-fold representation was the basis for determining its symbol in ref.\ \cite{Morales:2022csr}.
In this letter, we find similar differential equations and integral representations for the pentabox and double pentagon (Sec.\ \ref{sec:pbDE} and \ref{sec:dpDE}).
As part of the effort to fully understand planar two-loop scattering, we also study the singularity structure of these integrals, in particular generalising the simplicity found in the double-box symbol~\cite{Morales:2022csr} to the case of the remaining two-loop integrals, see Sec.\ \ref{sec:pbSym}.

\section{Differential equations and integral representations}
\label{sec: DEs and integral reps}

In this section, we describe the first-order differential equations satisfied by the pentabox and double-pentagon integrals, along with their resulting one-fold integral representations, which are similar to those of the double-box integral. To lay the groundwork, we begin by introducing our notations and reviewing the notable structures of the double-box integral, which will reappear in the pentabox and double-pentagon integrals, albeit in a slightly more intricate fashion.

\subsection{Review: Double box}\label{sec:dbDE}

\begin{figure}[t]
	\centering
	$ \raisebox{-7pt}{$\vcenter{\hbox{
	\begin{tikzpicture}[scale=0.54]
	%dual diagram
	\draw[mathyellow!50!white, thick] (-1.5,0) -- (1.5,0);
	\draw[mathyellow!50!white, thick] (0.5,-1.0) -- (0.5,1.0);
	\draw[mathyellow!50!white, thick] (-0.5,-1.0) -- (-0.5,1.0);
	%propagators
	\draw[mathblue, very thick] (0,-0.5)--(0,0.5);
	\draw[mathblue, very thick] (0,-0.5)--(1,-0.5);
	\draw[mathblue, very thick] (0,-0.5)--(-1,-0.5);
	\draw[mathblue, very thick] (0,0.5)--(1,0.5);
	\draw[mathblue, very thick] (0,0.5)--(-1,0.5);
	\draw[mathblue, very thick] (1,0.5)--(1,-0.5);
	\draw[mathblue, very thick] (-1,0.5)--(-1,-0.5);
	%external legs
	\draw[mathblue, very thick] (0,0.5) -- (-0.1,1.0);
	\draw[mathblue, very thick] (0,0.5) -- (0.1,1.0);
	\draw[mathblue, very thick] (0,-0.5) -- (-0.1,-1.0);
	\draw[mathblue, very thick] (0,-0.5) -- (0.1,-1.0);
	\draw[mathblue, very thick] (1,0.5) -- (1.42,0.78);
	\draw[mathblue, very thick] (1,0.5) -- (1.28,0.92);
	\draw[mathblue, very thick] (-1,0.5) -- (-1.42,0.78);
	\draw[mathblue, very thick] (-1,0.5) -- (-1.28,0.92);
	\draw[mathblue, very thick] (1,-0.5) -- (1.42,-0.78);
	\draw[mathblue, very thick] (1,-0.5) -- (1.28,-0.92);
	\draw[mathblue, very thick] (-1,-0.5) -- (-1.42,-0.78);
	\draw[mathblue, very thick] (-1,-0.5) -- (-1.28,-0.92);
	%labels
	\node[mathyellow!50!black] at (0.55,1.15) {\scriptsize $x_1$};
	\node[mathyellow!50!black] at (1.8,-0.05) {\scriptsize $x_2$};
	\node[mathyellow!50!black] at (0.55,-1.2) {\scriptsize $x_3$};
	\node[mathyellow!50!black] at (-0.5,-1.2) {\scriptsize $x_4$};
	\node[mathyellow!50!black] at (-1.75,-0.05) {\scriptsize $x_5$};
	\node[mathyellow!50!black] at (-0.5,1.15) {\scriptsize $x_6$};
	\node at (0,-1.8) {\small $(a)$};
	\end{tikzpicture}}}$}
	\xrightarrow[\text{eq.}\,\eqref{eq:dbDE}]{(1+\partial_L)}
	\raisebox{-7pt}{$
	\vcenter{\hbox{\begin{tikzpicture}[scale=0.54]
	%dual diagram
	\draw[mathyellow!50!white, thick] (-1.1,0) -- (1.1,0);
	\draw[mathyellow!50!white, thick] (-0.55,-0.95) -- (0.55,0.95);
	\draw[mathyellow!50!white, thick] (-0.55,0.95) -- (0.55,-0.95);
	%propagators
	\draw[mathblue, very thick] (-0.61,0.35)--(-0.61,-0.35);
	\draw[mathblue, very thick] (-0.61,-0.35) -- (0,-0.7);
	\draw[mathblue, very thick] (-0.61,0.35) -- (0,0.7);
	\draw[mathblue, very thick] (0.61,0.35)--(0.61,-0.35);
	\draw[mathblue, very thick] (0.61,0.35) -- (0,0.7);
	\draw[mathblue, very thick] (0.61,-0.35) -- (0,-0.7);
	%external legs
	\draw[mathblue, very thick] (0,0.7) -- (-0.1,1.2);
	\draw[mathblue, very thick] (0,0.7) -- (0.1,1.2);
	\draw[mathblue, very thick] (0,-0.7) -- (-0.1,-1.2);
	\draw[mathblue, very thick] (0,-0.7) -- (0.1,-1.2);
	\draw[mathblue, very thick] (0.61,0.35) -- (1.09,0.51);
	\draw[mathblue, very thick] (0.61,0.35) -- (0.99,0.68);
	\draw[mathblue, very thick] (-0.61,0.35) -- (-1.09,0.51);
	\draw[mathblue, very thick] (-0.61,0.35) -- (-0.99,0.68);
	\draw[mathblue, very thick] (-0.61,-0.35) -- (-1.09,-0.51);
	\draw[mathblue, very thick] (-0.61,-0.35) -- (-0.99,-0.68);
	\draw[mathblue, very thick] (0.61,-0.35) -- (1.09,-0.51);
	\draw[mathblue, very thick] (0.61,-0.35) -- (0.99,-0.68);
	%labels
	\node[mathyellow!50!black] at (0.7,1.1) {\scriptsize $x_1$};
	\node[mathyellow!50!black] at (1.4,-0.05) {\scriptsize $x_2$};
	\node[mathyellow!50!black] at (0.7,-1.1) {\scriptsize $x_3$};
	\node[mathyellow!50!black] at (-0.7,-1.1) {\scriptsize $x_4$};
	\node[mathyellow!50!black] at (-1.35,-0.05) {\scriptsize $x_5$};
	\node[mathyellow!50!black] at (-0.7,1.1) {\scriptsize $x_6$};
	\node at (0,-1.8) {\small $(b)$};
	\end{tikzpicture}}}$}$
	\caption{The four-dimensional double box $(a)$ and the related six-dimensional hexagon $(b)$, as well as their dual graphs. 
	}
	\label{fig:db}
	\end{figure}
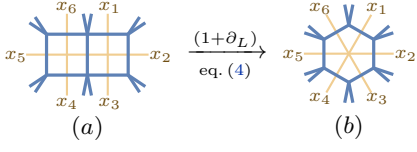

The most general double-box integral has six off-shell momenta $p_i^\mu\in\mathbb R^4$,
$i=1,...,6$, with $p_i^2\neq 0$. 
For planar massless Feynman integrals, it has proven useful to transition to the dual space with coordinates $x_i^\mu\in\mathbb R^4$ via $p_i=x_{i+1}-x_i$, cf.\ Fig.\ \ref{fig:db},
as this manifests a dual conformal symmetry of the associated Feynman integrals~\cite{Drummond:2006rz}. 
A conventional way to linearise this symmetry is to adopt the embedding-space formalism \cite{Mack:1969rr,Simmons-Duffin:2012juh}, where generally a point $x^{\mu}\in \mathbb{R}^{d}$ is mapped to a projective light ray $X^{M}=[1:x^{2}:x^{\mu}]\in\mathbb{R}^{d+1,1}$ (in light-cone coordinates), such that
\begin{equation}
	(x_{i}-x_{j})^2=-2X_{i} \cdot X_{j}=: (X_{i},X_{j})=:X_{ij} \,.
\end{equation}
In these coordinates, the double-box integral reads
\begin{align}
	\cI\big[\DB\big]^{\fourd}=
	\int\frac{D^4 X_0 \: D^4 X_{0^\prime}}{\left(\prod_{\ell \in L}X_{\ell 0}\right)X_{0{0^\prime}}\left(\prod_{r\in R}X_{r0^\prime}\right)}\,,
	\label{eq:dbIntegral}
\end{align}
where $R:= \{1,2,3\}$ and $L:=\{4,5,6\}$ are the index sets for the dual points $x_{i}$, cf.\ Fig.\ \ref{fig:db}. 
The integral measure $D^{d}X$ on the projective null cone is similarly defined as in ref.~\cite{Simmons-Duffin:2012juh}, see App.~\ref{app: Notations} for details.

A key driver of recent advances in the study of the double-box integral and in this letter is the differential equation~\cite{Paulos:2012nu,Nandan:2013ip}
\begin{align}
	\left(1+\partial_L\right)\cI\big[\DB\big]^{\fourd}=-\cI\big[\HEX\big]^\sixd \,,
	\label{eq:dbDE}
\end{align}
with $\partial_L:= {\textstyle\sum_{\{i,j\}\subset L}}X_{ij} \frac{\partial}{\partial{X_{ij}}}$, which relates the double box to the one-loop hexagon in six dimensions, a polylogarithm of weight three \cite{Ren:2023tuj}. 

The differential equation \eqref{eq:dbDE} can be recast into an integral relation.
In order to do so, we
 introduce a multiplicatively independent basis of dual-conformally invariant (DCI) cross ratios,
\begin{align}
	 \left\{\frac{X_{i,j-1}X_{i-1,j}}{X_{i,j}X_{i-1,j-1}}\bigg\vert~ \text{non-adjacent } \{i,j\} \subset [n] \right\}\,,
	\label{eq:chi}
\end{align}
with $[n]:=\{1,\ldots,n\}$. 
In the double-box case with $n=6$, there are nine such variables which we denote as $u_a$, $a\in[9]$, with respect to the lexicographical order of $\{i,j\}$.
For future convenience, we also introduce Gram determinants in embedding space as
\begin{equation}
	\mathcal{G}_{B}^{A}:=
	\det(X_{ab})_{\substack{a\in A\\ b\in B}}\ \ \ \text{and}\ \ \  \mathcal{G}_{A}:=\mathcal{G}_{A}^{A}\:,\quad A,B\subset [n]\,.
\end{equation}
They can be normalised by products of $X_{ij}$ to make them DCI, and we usually normalise such that they become \emph{polynomials} in the $u_{a}$'s;
for instance $\widehat{\mathcal{G}}_{[6]}:=\mathcal{G}_{[6]}/(X_{14}X_{25}X_{36})^2$, where the hat indicates normalisation.
Normalising both integrals in eq.\ \eqref{eq:dbDE} via $\widehat{\mathcal{I}}\big[\DB\big]^{\fourd}:=X_{13}X_{25}X_{46}\cI\big[\DB\big]^{\fourd}$ and $\widehat{\mathcal{I}}\big[\HEX\big]^\sixd:=\sqrt{-\mathcal{G}_{[6]}}\cI\big[\HEX\big]^\sixd$ renders them DCI and the hexagon integral becomes a pure polylogarithm.
The differential equation \eqref{eq:dbDE} yields 
\begin{align}
\widehat{\mathcal{I}}\big[\DB\big]^{\fourd}=\int_{u_{2}}^{\infty}\frac{dx}{y(x)}\widehat{\mathcal{I}}\big[\HEX\big]^\sixd(u_{2}\to x)
\label{eq:dbIE}
\end{align}
with $u_{2}\equiv X_{13}X_{46}/(X_{14} X_{36})$, cf.\ eq.\ \eqref{eq:chi}.
The fact that $y^{2}(x):=-\widehat{\mathcal G}_{[6]}(u_{2}\to x)$ is a polynomial of degree three in $x$ indicates the appearance of an elliptic curve in the double box. Nevertheless, it is as-yet-unknown whether the integral~\eqref{eq:dbIE} can be integrated to eMPLs.

Below we will see that the pentabox and double-pentagon integrals can similarly be expressed as (sums over) one-fold integrals of hexagons, with $u_2$ generalising to the respective cross ratios $u_2$ and $u_3$ of the four points around the \emph{middle} propagator, cf.\ Figs. \ref{fig: pb}(a) and \ref{fig: dp}(a).
To simplify the notation, we introduce $\widehat f(x):=\widehat f(u_{2(3)}\to x)$ for a DCI function $\widehat f$ where the corresponding cross ratio $u_2$ (or $u_3$) is replaced by the integration variable $x$.

\subsection{Pentabox}\label{sec:pbDE}

\begin{figure}[t] 
\begin{tikzpicture}[scale=0.54, baseline=-2pt]
%dual diagram
\draw[mathyellow!50!white, thick] (-0.7,0) -- (1.5,0);
\draw[mathyellow!50!white, thick] (0.5,-1.0) -- (0.5,1.0);
\draw[mathyellow!50!white, thick] (-0.7,0) -- (-0.4,1.1);
\draw[mathyellow!50!white, thick] (-0.7,0) -- (-0.4,-1.1);
\draw[mathyellow!50!white, thick] (-0.7,0) -- (-1.65,0.7);
\draw[mathyellow!50!white, thick] (-0.7,0) -- (-1.65,-0.7);
%propagators
\draw[mathblue, very thick] (0,-0.5)--(0,0.5);
\draw[mathblue, very thick] (0,-0.5)--(-0.95,-0.8);
\draw[mathblue, very thick] (0,0.5)--(-0.95,0.8);
\draw[mathblue, very thick] (0,-0.5)--(1,-0.5);
\draw[mathblue, very thick] (0,0.5)--(1,0.5);
\draw[mathblue, very thick] (-1.55,0)--(-0.95,-0.8);
\draw[mathblue, very thick] (-1.55,0)--(-0.95,0.8);
\draw[mathblue, very thick] (1,0.5)--(1,-0.5);
%external legs
\draw[mathblue, very thick] (0,0.5) -- (-0.1,1.0);
\draw[mathblue, very thick] (0,0.5) -- (0.1,1.0);
\draw[mathblue, very thick] (0,-0.5) -- (-0.1,-1.0);
\draw[mathblue, very thick] (0,-0.5) -- (0.1,-1.0);
\draw[mathblue, very thick] (-1.55,0) -- (-2.05,0.1);
\draw[mathblue, very thick] (-1.55,0) -- (-2.05,-0.1);
\draw[mathblue, very thick] (1,0.5) -- (1.42,0.78);
\draw[mathblue, very thick] (1,0.5) -- (1.28,0.92);
\draw[mathblue, very thick] (1,-0.5) -- (1.42,-0.78);
\draw[mathblue, very thick] (1,-0.5) -- (1.28,-0.92);
\draw[mathblue, very thick] (-0.95,-0.8) -- (-1.20,-1.25);
\draw[mathblue, very thick] (-0.95,-0.8) -- (-1.00,-1.31);
\draw[mathblue, very thick] (-0.95,0.8) -- (-1.20,1.25);
\draw[mathblue, very thick] (-0.95,0.8) -- (-1.00,1.31);
%labels
\node[mathyellow!50!black] at (0.55,1.15) {\scriptsize $x_1$};
\node[mathyellow!50!black] at (1.8,-0.05) {\scriptsize $x_2$};
\node[mathyellow!50!black] at (0.55,-1.2) {\scriptsize $x_3$};
\node[mathyellow!50!black] at (-0.3,-1.3) {\scriptsize $x_4$};
\node[mathyellow!50!black] at (-1.65,-0.9) {\scriptsize $x_5$};
\node[mathyellow!50!black] at (-1.65,0.87) {\scriptsize $x_6$};
\node[mathyellow!50!black] at (-0.3,1.25) {\scriptsize $x_7$};
\node[mathblue!50!black] at (-0.7,0) {$\numpb$};
\node at (0,-2) {\small $(a)$};
\end{tikzpicture}
\hspace{-5pt}$\xleftarrow[\text{eq.}\,\eqref{eq: pb_reduction}]{\sum_{r}(N,X_r)}$
\begin{tikzpicture}[scale=0.54, baseline=-2pt]
%dual diagram
\draw[mathyellow!50!white, thick] (-0.7,0) -- (1.5,0);
\draw[mathyellow!50!white, thick] (0.5,-1.0) -- (0.5,1.0);
\draw[mathyellow!50!white, thick] (-0.7,0) -- (-0.4,1.1);
\draw[mathyellow!50!white, thick] (-0.7,0) -- (-0.4,-1.1);
\draw[mathyellow!50!white, thick] (-0.7,0) -- (-1.65,0.7);
\draw[mathyellow!50!white, thick] (-0.7,0) -- (-1.65,-0.7);
%propagators
\draw[mathblue, very thick] (0,-0.5)--(0,0.5);
\draw[mathblue, very thick] (0,-0.5)--(-0.95,-0.8);
\draw[mathblue, very thick] (0,0.5)--(-0.95,0.8);
\draw[mathblue, very thick] (0,-0.5)--(1,-0.5);
\draw[mathblue, very thick] (0,0.5)--(1,0.5);
\draw[mathblue, very thick] (-1.55,0)--(-0.95,-0.8);
\draw[mathblue, very thick] (-1.55,0)--(-0.95,0.8);
\draw[mathblue, very thick] (1,0.5)--(1,-0.5);
\fill[mathblue] (0,0) circle (0.1);
\fill[mathblue] (0.5,-0.5) circle (0.1);
%external legs
\draw[mathblue, very thick] (0,0.5) -- (-0.1,1.0);
\draw[mathblue, very thick] (0,0.5) -- (0.1,1.0);
\draw[mathblue, very thick] (0,-0.5) -- (-0.1,-1.0);
\draw[mathblue, very thick] (0,-0.5) -- (0.1,-1.0);
\draw[mathblue, very thick] (-1.55,0) -- (-2.05,0.1);
\draw[mathblue, very thick] (-1.55,0) -- (-2.05,-0.1);
\draw[mathblue, very thick] (1,0.5) -- (1.42,0.78);
\draw[mathblue, very thick] (1,0.5) -- (1.28,0.92);
\draw[mathblue, very thick] (1,-0.5) -- (1.42,-0.78);
\draw[mathblue, very thick] (1,-0.5) -- (1.28,-0.92);
\draw[mathblue, very thick] (-0.95,-0.8) -- (-1.20,-1.25);
\draw[mathblue, very thick] (-0.95,-0.8) -- (-1.00,-1.31);
\draw[mathblue, very thick] (-0.95,0.8) -- (-1.20,1.25);
\draw[mathblue, very thick] (-0.95,0.8) -- (-1.00,1.31);
%labels
\node at (0,-2) {\small $(b)$};
\end{tikzpicture}
$\xrightarrow[\text{eq.}\,\eqref{eq:pbDE}]{(1+\partial_L)}$
\begin{tikzpicture}[scale=0.54, baseline=-2pt]
%dual diagram
\draw[mathyellow!50!white, thick] (0,0) -- (1.3,0);
\draw[mathyellow!50!white, thick] (0,0) -- (0.87,1.02);
\draw[mathyellow!50!white, thick] (0,0) -- (0.87,-1.02);
\draw[mathyellow!50!white, thick] (0,0) -- (-0.23,1.27);
\draw[mathyellow!50!white, thick] (0,0) -- (-0.23,-1.27);
\draw[mathyellow!50!white, thick] (0,0) -- (-1.17,0.56);
\draw[mathyellow!50!white, thick] (0,0) -- (-1.17,-0.56);
%propagators
\draw[mathblue, very thick] (0.77,-0.35)--(0.77,0.35);
\draw[mathblue, very thick] (0.77,-0.35)--(0.22,-0.79);
\draw[mathblue, very thick] (0.77,0.35)--(0.22,0.79);
\draw[mathblue, very thick] (0.22,0.79) -- (-0.46,0.63);
\draw[mathblue, very thick] (0.22,-0.79) -- (-0.46,-0.63);
\draw[mathblue, very thick] (-0.46,-0.63) -- (-0.77,0);
\draw[mathblue, very thick] (-0.46,0.63) -- (-0.77,0);
\fill[mathblue] (0.49,-0.57) circle (0.1);
%external legs
\draw[mathblue, very thick] (-0.77,0) -- (-1.28,0.1);
\draw[mathblue, very thick] (-0.77,0) -- (-1.28,-0.1);
\draw[mathblue, very thick] (-0.46,-0.63) -- (-0.85,-0.96);
\draw[mathblue, very thick] (-0.46,-0.63) -- (-0.69,-1.08);
\draw[mathblue, very thick] (-0.46,0.63) -- (-0.85,0.96);
\draw[mathblue, very thick] (-0.46,0.63) -- (-0.69,1.08);
\draw[mathblue, very thick] (0.22,-0.79) -- (0.23,-1.30);
\draw[mathblue, very thick] (0.22,-0.79) -- (0.43,-1.26);
\draw[mathblue, very thick] (0.22,0.79) -- (0.23,1.30);
\draw[mathblue, very thick] (0.22,0.79) -- (0.43,1.26);
\draw[mathblue, very thick] (0.77,-0.35) -- (1.18,-0.66);
\draw[mathblue, very thick] (0.77,-0.35) -- (1.26,-0.48);
\draw[mathblue, very thick] (0.77,0.35) -- (1.18,0.66);
\draw[mathblue, very thick] (0.77,0.35) -- (1.26,0.48);
%labels
\node[mathyellow!50!black] at (1.15,-1.15) {\scriptsize $x_3$};
\node at (0,-2) {\small $(c)$};
\end{tikzpicture}
\caption{The four-dimensional pentabox $(a)$ and the related six-dimensional pentabox $(b)$ and eight-dimensional heptagon $(c)$, as well as their dual graphs.}
\label{fig: pb}
\end{figure}

The pentabox integral in the basis of planar two-loop massless Feynman integrals \eqref{eq: generic two loop} is 
\begin{align}
\label{eq: tensor pentabox}
	\cI\big[\PB\big]^{\fourd}=
	\int\frac{D^4 X_0 D^4 X_{0'}~(N,X_0)}{\left(\prod_{\ell\in L}X_{\ell 0}\right) X_{00^\prime}\left(\prod_{r\in R}X_{r0^\prime}\right)}\,,
\end{align}
where $N\in\mathbb R^{5,1}$ is orthogonal to all $X_{\ell}$ for $\ell \in L$, and now $R:=\{1,2,3\}$ and $L:=\{4,5,6,7\}$, cf.\ Fig.\ \ref{fig: pb}(a). 

We find that the pentabox tensor integral \eqref{eq: tensor pentabox} in four dimensions can be related to scalar pentaboxes in six dimensions via
\begin{align} \label{eq: pb_reduction}
	\cI\big[\PB\big]^\fourd=&~(N,X_1)\,\cI\big[\PBone\big]^\sixd 
	+(N,X_2)\,\cI\big[\PBtwo\big]^\sixd\nonumber\\
	&+(N,X_3)\,\cI\big[\PBthree\big]^\sixd \,,
\end{align}
where lines with a dot 
 represent the corresponding propagators being squared. 
Each of these six-dimensional pentaboxes is related to an eight-dimensional  heptagon integral with one doubled propagator through a generalisation of eq.\ \eqref{eq:dbDE}, e.g.
\begin{align} \label{eq:pbDE}
(1+\partial_L)\cI\big[\PBthree\big]^\sixd\!\!&=-\cI\big[\HEP\big]^\eightd~ \!\!\!  \\ 
&=-\sum_{i=1}^{7}\frac{(-1)^{i+3}\cG_{[\hat{3}]}^{[\hat{\imath}]}}{\cG_{[7]}} \cI \big[\HEX\big]^{\sixd}_{[\hat{\imath}]}\,, \label{eq:hep_to_hex}%\nonumber
\end{align}
where $[\hat \imath]:=[7]\setminus\{i\}$, and in the second relation we have applied the one-loop reduction of ref.~\cite{Arkani-Hamed:2017ahv} to reduce the heptagon to a sum of six-dimensional hexagons, with subscript $[\hat{\imath}]$ indicating the index set of the six dual points. 

Relations \eqref{eq: pb_reduction} and \eqref{eq:pbDE} can be easily verified in the Mellin-space framework of refs.\ \cite{Paulos:2012nu,Nandan:2013ip}, or via Feynman-parameter representations obtained by loop-by-loop integration~\cite{Bourjaily:2019jrk}. 
A subtle point to highlight is that in deriving eq.~\eqref{eq:pbDE} we assumed all $X_{ij}$'s to be independent.
However, these $X_{ij}$'s are subject to a constraint, $\mathcal{G}_{[7]}=0$, in four-dimensional kinematics. 
At the level of the 14 cross ratios of eq.~\eqref{eq:chi} (with $n=7$), this becomes the constraint $\widehat{\mathcal{G}}_{[7]}=0$, which reduces the number of independent kinematic degrees of freedom to 13 and implies
\begin{equation} \label{eq: constraint_4d}
	\widehat{\mathcal{G}}_{[7]}(x)\propto (x-u_{2})(x-r_{+})(x-r_{-}) \,,
\end{equation}
where now $u_{2}:=X_{13}X_{47}/(X_{14} X_{37})$ and the replacement $u_2\to x$ is understood to occur before restricting to four dimensions.
The constraint $\cG_{[7]}=0$ does not commute with the differential operator in eq.\ \eqref{eq:pbDE} and thus we first solve this equation for generic $X_{ij}$'s, resulting in a one-fold integral representation over hexagon integrals (via eq.\ \eqref{eq:hep_to_hex}) for the six-dimensional pentaboxes, and afterwards restrict to four-dimensional external kinematics.

Together with eq.\ \eqref{eq: pb_reduction}, eqs.\ \eqref{eq:pbDE}--\eqref{eq:hep_to_hex} lead to a one-fold integral representation over hexagon integrals:
\begin{equation}
\label{eq: one-fold over hexagon integrals with LS}
	\cI\big[\PB\big]^{\fourd}= \sum_{\sigma=\pm}\sf{LS}_{\text{pb}}^{(\sigma)}~ 
	\widehat{\cI}\big[\PBnoN\big]^{(\sigma)}\,,
\end{equation}
where $\sf{LS}^{(\pm)}_{\text{pb}}$ is the sum/difference of the two leading singularities of the pentabox (see App.~\ref{app: Notations} for explicit expressions), and\footnote{Accordingly, the labelling of $r_{\pm}$ is such that $r_{+}-r_{-}$ is the positive square root of the discriminant of the quadratic polynomial $\widehat{\cG}_{[7]}(x)/(x-u_{2})$.} 
\begin{equation} \label{eq: int_rep_pb}
	\widehat{\cI}\big[\PBnoN\big]^{(\pm)}=\sum_{i=1}^{7}(-1)^{i} \int^{\infty}_{u_{2}}\frac{dx~y_{i}(r_{\pm})}{y_{i}(x)(x-r_{\pm})}~ \widehat{\cI}[\HEX]^{\sixd}_{[\hat{\imath}]}(x)\,.
\end{equation}
Here $y_{i}^{2}(x):=-\widehat{\cG}_{[\hat{\imath}]}(x)$ is 
cubic in $x$ only for $i\in L$ (and otherwise quadratic); thus, eq.\ \eqref{eq: int_rep_pb} contains four elliptic curves, corresponding to the four double-box subtopologies of the pentabox.
Furthermore, note that the integral in eq.\ \eqref{eq: int_rep_pb} appears to be singular as $u_{2}\to r_{\pm}$, seemingly conflicting with the \emph{first-entry} condition~\cite{Gaiotto:2011dt}, which states that massless planar Feynman integrals can only become singular at $x_{ij}^{2}=0$. This apparent contradiction is resolved by the relation
\begin{equation} 
	\sum_{i=1}^{7} (-1)^{i}~\widehat{\cI}[\HEX]_{[\hat{\imath}]}^{\sixd}(x\to r_{\pm})=0 \,,
	\label{eq:sing_cancel}
\end{equation}
which is due to the interpretation of one-loop integrals as volumes of simplices on hyperbolic manifolds~\cite{Davydychev:1997wa,Zagier2007}. 
We will see below that similar cancellations occur at the symbol level.

\subsection{Double pentagon}\label{sec:dpDE}

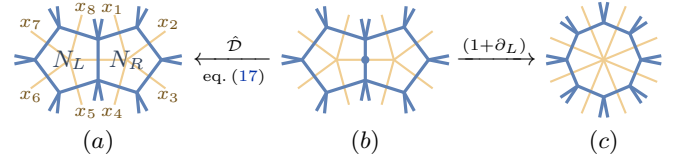
\begin{figure}[t]
	\begin{tikzpicture}[scale=0.54,baseline=-2pt]
	%dual diagram
	\draw[mathyellow!50!white, thick] (-0.7,0) -- (0.7,0);
	\draw[mathyellow!50!white, thick] (-0.7,0) -- (-0.4,1.1);
	\draw[mathyellow!50!white, thick] (0.7,0) -- (0.4,1.1);
	\draw[mathyellow!50!white, thick] (-0.7,0) -- (-0.4,-1.1);
	\draw[mathyellow!50!white, thick] (0.7,0) -- (0.4,-1.1);
	\draw[mathyellow!50!white, thick] (-0.7,0) -- (-1.65,0.7);
	\draw[mathyellow!50!white, thick] (0.7,0) -- (1.65,0.7);
	\draw[mathyellow!50!white, thick] (-0.7,0) -- (-1.65,-0.7);
	\draw[mathyellow!50!white, thick] (0.7,0) -- (1.65,-0.7);
	%propagators
	\draw[mathblue, very thick] (0,-0.5)--(0,0.5);
	\draw[mathblue, very thick] (0,-0.5)--(-0.95,-0.8);
	\draw[mathblue, very thick] (0,-0.5)--(0.95,-0.8);
	\draw[mathblue, very thick] (0,0.5)--(-0.95,0.8);
	\draw[mathblue, very thick] (0,0.5)--(0.95,0.8);
	\draw[mathblue, very thick] (-1.55,0)--(-0.95,-0.8);
	\draw[mathblue, very thick] (1.55,0)--(0.95,-0.8);
	\draw[mathblue, very thick] (-1.55,0)--(-0.95,0.8);
	\draw[mathblue, very thick] (1.55,0)--(0.95,0.8);
	%external legs
	\draw[mathblue, very thick] (0,0.5) -- (-0.1,1.0);
	\draw[mathblue, very thick] (0,0.5) -- (0.1,1.0);
	\draw[mathblue, very thick] (0,-0.5) -- (-0.1,-1.0);
	\draw[mathblue, very thick] (0,-0.5) -- (0.1,-1.0);
	\draw[mathblue, very thick] (-1.55,0) -- (-2.05,0.1);
	\draw[mathblue, very thick] (-1.55,0) -- (-2.05,-0.1);
	\draw[mathblue, very thick] (1.55,0) -- (2.05,0.1);
	\draw[mathblue, very thick] (1.55,0) -- (2.05,-0.1);
	\draw[mathblue, very thick] (-0.95,-0.8) -- (-1.20,-1.25);
	\draw[mathblue, very thick] (-0.95,-0.8) -- (-1.00,-1.31);
	\draw[mathblue, very thick] (0.95,-0.8) -- (1.20,-1.25);
	\draw[mathblue, very thick] (0.95,-0.8) -- (1.00,-1.31);
	\draw[mathblue, very thick] (-0.95,0.8) -- (-1.20,1.25);
	\draw[mathblue, very thick] (-0.95,0.8) -- (-1.00,1.31);
	\draw[mathblue, very thick] (0.95,0.8) -- (1.20,1.25);
	\draw[mathblue, very thick] (0.95,0.8) -- (1.00,1.31);
	%labels
	\node[mathyellow!50!black] at (0.35,1.24) {\scriptsize $x_1$};
	\node[mathyellow!50!black] at (1.72,0.87) {\scriptsize $x_2$};
	\node[mathyellow!50!black] at (1.72,-0.9) {\scriptsize $x_3$};
	\node[mathyellow!50!black] at (0.35,-1.3) {\scriptsize $x_4$};
	\node[mathyellow!50!black] at (-0.3,-1.3) {\scriptsize $x_5$};
	\node[mathyellow!50!black] at (-1.65,-0.9) {\scriptsize $x_6$};
	\node[mathyellow!50!black] at (-1.65,0.87) {\scriptsize $x_7$};
	\node[mathyellow!50!black] at (-0.3,1.25) {\scriptsize $x_8$};
	\node[mathblue!50!black] at (-0.7,0) {$\numdpl$};
	\node[mathblue!50!black] at (0.7,0) {$\numdpr$};
	\node at (0,-2) {\small $(a)$};
	\end{tikzpicture}
	\hspace{-1pt}$\xleftarrow[\text{eq.}\,\eqref{eq:dpDO}]{\hat{\mathcal D}}$
	\begin{tikzpicture}[scale=0.54,baseline=-2pt]
	%dual diagram
	\draw[mathyellow!50!white, thick] (-0.7,0) -- (0.7,0);
	\draw[mathyellow!50!white, thick] (-0.7,0) -- (-0.4,1.1);
	\draw[mathyellow!50!white, thick] (0.7,0) -- (0.4,1.1);
	\draw[mathyellow!50!white, thick] (-0.7,0) -- (-0.4,-1.1);
	\draw[mathyellow!50!white, thick] (0.7,0) -- (0.4,-1.1);
	\draw[mathyellow!50!white, thick] (-0.7,0) -- (-1.65,0.7);
	\draw[mathyellow!50!white, thick] (0.7,0) -- (1.65,0.7);
	\draw[mathyellow!50!white, thick] (-0.7,0) -- (-1.65,-0.7);
	\draw[mathyellow!50!white, thick] (0.7,0) -- (1.65,-0.7);
	%propagators
	\draw[mathblue, very thick] (0,-0.5)--(0,0.5);
	\draw[mathblue, very thick] (0,-0.5)--(-0.95,-0.8);
	\draw[mathblue, very thick] (0,-0.5)--(0.95,-0.8);
	\draw[mathblue, very thick] (0,0.5)--(-0.95,0.8);
	\draw[mathblue, very thick] (0,0.5)--(0.95,0.8);
	\draw[mathblue, very thick] (-1.55,0)--(-0.95,-0.8);
	\draw[mathblue, very thick] (1.55,0)--(0.95,-0.8);
	\draw[mathblue, very thick] (-1.55,0)--(-0.95,0.8);
	\draw[mathblue, very thick] (1.55,0)--(0.95,0.8);
	%external legs
	\draw[mathblue, very thick] (0,0.5) -- (-0.1,1.0);
	\draw[mathblue, very thick] (0,0.5) -- (0.1,1.0);
	\draw[mathblue, very thick] (0,-0.5) -- (-0.1,-1.0);
	\draw[mathblue, very thick] (0,-0.5) -- (0.1,-1.0);
	\draw[mathblue, very thick] (-1.55,0) -- (-2.05,0.1);
	\draw[mathblue, very thick] (-1.55,0) -- (-2.05,-0.1);
	\draw[mathblue, very thick] (1.55,0) -- (2.05,0.1);
	\draw[mathblue, very thick] (1.55,0) -- (2.05,-0.1);
	\draw[mathblue, very thick] (-0.95,-0.8) -- (-1.20,-1.25);
	\draw[mathblue, very thick] (-0.95,-0.8) -- (-1.00,-1.31);
	\draw[mathblue, very thick] (0.95,-0.8) -- (1.20,-1.25);
	\draw[mathblue, very thick] (0.95,-0.8) -- (1.00,-1.31);
	\draw[mathblue, very thick] (-0.95,0.8) -- (-1.20,1.25);
	\draw[mathblue, very thick] (-0.95,0.8) -- (-1.00,1.31);
	\draw[mathblue, very thick] (0.95,0.8) -- (1.20,1.25);
	\draw[mathblue, very thick] (0.95,0.8) -- (1.00,1.31);
	%labels
	\fill[mathblue] (0,0) circle (0.1);
	\node at (0,-2) {\small $(b)$};
	\end{tikzpicture}
	\hspace{-1pt}$\xrightarrow[]{(1+\partial_L)}$
	\begin{tikzpicture}[scale=0.54,baseline=-2pt]
	%dual diagram
	\draw[mathyellow!50!white, thick] (-1.31,-0.54) -- (1.31,0.54);
	\draw[mathyellow!50!white, thick] (1.31,-0.54) -- (-1.31,0.54);
	\draw[mathyellow!50!white, thick] (-0.54,-1.31) -- (0.54,1.31);
	\draw[mathyellow!50!white, thick] (0.54,-1.31) -- (-0.54,1.31);
	%propagators
	\draw[mathblue, very thick] (-0.91,0)--(-0.64,-0.64);
	\draw[mathblue, very thick] (0.64,0.64)--(0.91,0);
	\draw[mathblue, very thick] (-0.64,-0.64)--(0,-0.91);
	\draw[mathblue, very thick] (0.64,-0.64)--(0,-0.91);
	\draw[mathblue, very thick] (-0.64,0.64)--(0,0.91);
	\draw[mathblue, very thick] (0.64,0.64)--(0,0.91);
	\draw[mathblue, very thick] (-0.91,0)--(-0.64,0.64);
	\draw[mathblue, very thick] (0.91,0)--(0.64,-0.64);
	%external legs
	\draw[mathblue, very thick] (0,0.91) -- (0.1,1.41);
	\draw[mathblue, very thick] (0,0.91) -- (-0.1,1.41);
	\draw[mathblue, very thick] (0,-0.91) -- (0.1,-1.41);
	\draw[mathblue, very thick] (0,-0.91) -- (-0.1,-1.41);
	\draw[mathblue, very thick] (0.91,0) -- (1.41,0.1);
	\draw[mathblue, very thick] (0.91,0) -- (1.41,-0.1);
	\draw[mathblue, very thick] (-0.91,0) -- (-1.41,0.1);
	\draw[mathblue, very thick] (-0.91,0) -- (-1.41,-0.1);
	\draw[mathblue, very thick] (0.64,0.64) -- (1.06,0.92);
	\draw[mathblue, very thick] (0.64,0.64) -- (0.92,1.06);
	\draw[mathblue, very thick] (-0.64,0.64) -- (-1.06,0.92);
	\draw[mathblue, very thick] (-0.64,0.64) -- (-0.92,1.06);
	\draw[mathblue, very thick] (0.64,-0.64) -- (1.06,-0.92);
	\draw[mathblue, very thick] (0.64,-0.64) -- (0.92,-1.06);
	\draw[mathblue, very thick] (-0.64,-0.64) -- (-1.06,-0.92);
	\draw[mathblue, very thick] (-0.64,-0.64) -- (-0.92,-1.06);
	%labels
	\node at (0,-2) {\small $(c)$};
	\end{tikzpicture}
	\caption{The four-dimensional double pentagon $(a)$ and the related six-dimensional double pentagon $(b)$ and eight-dimensional octagon $(c)$, as well as their dual graphs.} 
	\label{fig: dp}
	\end{figure}

Now, let us turn to the double pentagon,
\begin{align}
\cI\big[\DP\big]^{\fourd}=
\int\frac{D^4 X_0 D^4 X_{0^\prime}~(N_L,X_0)(N_R, X_{0^\prime})}{\left(\prod_{\ell\in L}X_{\ell 0}\right)
X_{00^\prime}\left(\prod_{r\in R}X_{r0^\prime}\right)},\,
\end{align}
where $N_{L(R)}\in\mathbb R^{5,1}$ is orthogonal to $X_{i}$ for all $i\in L(R)$ with $R:=\{1,2,3,4\}$ and $L:=\{5,6,7,8\}$, cf.\ Fig.\ \ref{fig: dp}(a).
This integral depends on 17 independent cross ratios in four-dimensional external kinematics.

The story here is analogous to the pentabox case: the double pentagon in four dimensions can be obtained from the six-dimensional scalar double pentagon with the middle propagator being doubled by acting with the operator
\begin{equation}
	\hat{\mathcal{D}}:=-(N_L,N_R)-\!\!\!\!\sum_{\ell\in L,r\in R}\!\!\!\!(N_L,X_r)(N_R,X_\ell)\frac{\partial}{\partial X_{\ell r}} \,,
\label{eq:dpDO}
\end{equation}
while the scalar double pentagon is related to the octagon in eight dimensions via $(1+\partial_{L})$, cf.\ Fig.~\ref{fig: dp}.

After some algebra, one can show from these relations that also the double-pentagon integral allows for a one-fold integral representation over hexagon integrals:
\begin{align}\label{eq: int_rep_dp}
&\cI\big[\DP\big]^{\fourd}=
	\sum_{i=1}^8\sum_{\sigma=\pm} (-1)^{i+1}~ \mathsf{LS}_{\text{pb}[\hat\imath]}^{(\sigma)}~ \widehat{\cI}\big[\PBnoN\big]^{(\sigma)}_{[\hat\imath]}\\
&-\!\!\!\sum_{\ell\in L, r\in R}\!\!\!(-1)^{\ell +r}\!\! \int^{\infty}_{u_{3}}\!\! dx \biggl(\frac{\mathsf{LS}_{\text{kb}}\,y_{\ell r}(0)}{x\: y_{\ell r}(x)}-
	\frac{C_{\ell r}}{y_{\ell r}(x)}\biggr) \widehat{\cI} \big[ \HEX \big]^{\sixd}_{[\hat{\ell}\hat{r}]}\!(x)\,.\nonumber
\end{align}
In the first term on the RHS, the integrals \eqref{eq: int_rep_pb} reappear, with $[\hat\imath]:=[8]\setminus \{i\}$ indicating the relevant pentabox index sets, together with the prefactor $\mathsf{LS}_{\text{pb}[\hat\imath]}^{(\pm)}$ corresponding to the sum/difference of the double-pentagon residues on the associated pentabox cut. In the second term, $[\hat{\ell}\hat{r}]:=[8]\setminus \{\ell, r\}$ indicates the index set for the hexagon integrals and $y_{\ell r}^2(x):=-\widehat{\cG}_{[\hat{\ell}\hat{r}]}(x)$ is cubic in $x$ in all cases, i.e.\ there appear 16 elliptic curves in this expression, corresponding to the 16 double-box subtopologies of the double pentagon. Finally, $\mathsf{LS}_{\text{kb}}$ 
 is a linear combination of residues on the kissing-box cut and the $C_{\ell r}$'s 
are rational functions of the kinematics. See App.~\ref{app: Notations} for more explicit expressions.

\section{Symbol}\label{sec:dbSym}
\label{sec:pbSym}
\label{sec:dpSym}

In this section, we study the symbols and coproducts of the pentabox and double-pentagon integral. 
For a general $n$-fold MPL or eMPL $I_n$, the symbol can be defined through the total differential $d I_{n} = \sum_{\alpha} I_{n-1}^{(\alpha)} d f_{\alpha}$, where $\{f_\alpha\}$ is a set of one-fold integrals:
\begin{equation}
\label{eq: def symbol}
	\mathcal{S}(I_{n}) = \textstyle\sum_{\alpha} \mathcal{S}(I_{n-1}^{(\alpha)})\otimes f_{\alpha}\,.
\end{equation}
Although some information is lost in this mapping, it uncovers key algebraic properties of the original integral, such as functional identities and its singularity structure~\cite{Goncharov:2009lql,Goncharov:2010jf,Duhr:2011zq,Duhr:2012fh}.
The elliptic symbol bootstrap \cite{Morales:2022csr} and the algorithm developed in ref.\ \cite{He:2023qld} have in some cases been successful at computing symbols of elliptic Feynman integrals when their explicit forms in terms of eMPLs are unknown.
However, leveraging the structure of the double-box symbol, the one-fold integral representations from the previous section, and further known properties of (elliptic) symbols, we will see that the pentabox and double-pentagon symbols are highly constrained, essentially allowing us to write down the answer directly!

The symbol of the double-box integral \cite{Morales:2022csr} can be organised into a compact $\Delta_{2,2}$ coproduct, defined in a similar way to eq.\ \eqref{eq: def symbol}.\footnote{Roughly speaking, the coproduct can be thought of as a partial uplift of the symbol to the full function; for instance, eq.~\eqref{eq: def symbol} can be understand as $\Delta_{n-1,1}(I_{n}) = \textstyle\sum_{\alpha} I_{n-1}^{(\alpha)}\otimes f_{\alpha}$ in terms of $\Delta_{n-1,1}$ coproducts. See refs.~\cite{Gonch2,Goncharov:2001iea,Duhr:2012fh} for further details.} Specifically,
\begin{align} \label{eq: coproduct_db}
	&\Delta_{2,2}\Big({-}\frac{2\pi i}{\omega_1}\widehat{\cI}\big[\DB\big]^\fourd\Big)\\[-12pt]
	&=\sum_{i<j}\widehat{\cI}\big[\SB\big]_{[\hat \imath\hat \jmath]}^\fourd\otimes \overbrace{\frac{2\pi i}{\omega_{1}}\int^{u_{2}}\!\!\!\frac{dx}{y(x)}  \log R_{ij}(x)}^{\qquad =:S_{ij}} - (u_{2}\to\infty)\,,\nonumber
\end{align}
where $\omega_1$ and $\omega_2$ are the periods of the elliptic curve, $\widehat{\cI}\big[\SB\big]^\fourd_{[\hat{\imath}\hat{\jmath}]}$ is the four-mass box integral with $[\hat \imath\hat \jmath]$ indicating the respective dual points, and $\log R_{ij}$ are the last entries of the hexagon symbol \cite{Spradlin:2011wp}, 
\begin{align}\label{eq:hex_symb}
R_{ij} :=\frac{(-1)^{i+j}\widehat{\cG}^{[\hat{\imath}]}_{[\hat{\jmath}]}-
	\sqrt{-\widehat{\cG}_{[6]}\widehat{\cG}_{[\hat{\imath}\hat{\jmath}]}}}{(-1)^{i+j}\widehat{\cG}^{[\hat{\imath}]}_{[\hat{\jmath}]}+
	\sqrt{-\widehat{\cG}_{[6]}\widehat{\cG}_{[\hat{\imath}\hat{\jmath}]}}} \,.
\end{align}
The symbols of $\widehat{\cI}\big[\SB\big]^\fourd_{[\hat{\imath}\hat{\jmath}]}$ and $S_{ij}$ are provided in App.~\ref{app: symbs}. The presence of only four-mass box integrals in the first entry of eq.\ \eqref{eq: coproduct_db} manifests the \emph{first-two-entries condition}~\cite{Gaiotto:2011dt,Caron-Huot:2011zgw,He:2021mme,Morales:2022csr}.
Note that the structure \eqref{eq: coproduct_db} mimics that of the hexagon coproduct 
 \cite{Spradlin:2011wp}
\begin{equation}\label{eq:hex_coproduct}
	\Delta_{2,1}\left(\widehat{\cI} \big[ \HEX \big]^\sixd\right)=\sum_{i<j}\widehat{\cI}\big[\SB\big]^\fourd_{[\hat{\imath}\hat{\jmath}]}\otimes \log R_{ij}\,,
\end{equation}
and the one-fold integral representation~\eqref{eq:dbIE}.

A new ingredient when going from the double box to pentabox and double pentagon are integrals of the form
\begin{equation} \label{eq: new_general_int}
	\tilde{\mathcal{I}}(u,c):=\int_{u}^{\infty}\frac{dx~ y(c)}{y(x)(x-c)} \: \widehat{\cI} \big[\HEX\big]^{\sixd}(x)
\end{equation}
for a generic elliptic curve $y(x)$, cf.\ eqs.\ \eqref{eq: int_rep_pb} and \eqref{eq: int_rep_dp}.
A technical but crucial detail is that the integration kernel in eq.~\eqref{eq: new_general_int} is a combination of two pure integration kernels defined on the torus $\mathbb C^2/(\mathbb{Z}+\mathbb{Z}\tau)$ associated to the elliptic curve, with modular parameter $\tau=\omega_{2}/\omega_{1}$. 
Specifically,
\begin{align}\label{eq:kernels}
\frac{ dx~ y(c) }{y(x)(x{-}c)}= dw(x)\Bigl(
	\partial_{z}\tilde\Omega(z,w(c))\big|_{z=w(x)}\!  -
	\partial_{z}\tilde\Omega(z,w(c))\big|_{z=0}\Bigr)
\end{align}
with the torus coordinate $w({\bullet}):=\omega^{-1}_{1}\int^{\bullet} dx/y(x)$,
and $\tilde\Omega(z,w):=\log \frac{\theta_{1}(z-w|\tau)}{\theta_{1}(z+w|\tau)}$, where $\theta_{1}$ denotes the odd Jacobi theta function. 
In the notation of ref.~\cite{Broedel:2018qkq}, the first integration kernel on the RHS of eq.\ \eqref{eq:kernels} corresponds to $dx\Psi_{-1}(c,x)$ and the latter is related to the double-box integration kernel $dx/y(x)$. 
To focus only on the new structure, we introduce the map $\mathscr I$,
\begin{equation} \label{eq: general integral}
	\mathscr I[\tilde{\mathcal{I}}]:=\int_{u}^{\infty}\!\!dx\,\Psi_{-1}(c,x) \: \widehat{\cI} \big[\HEX\big]^{\sixd}(x)\,,
\end{equation}
which removes double-box-like terms from $\tilde{\mathcal{I}}$.\footnote{The pentabox representation \eqref{eq: int_rep_pb} contains next to the elliptic cases ($i\in L$) also polylogarithmic integrals ($i\in R$), i.e.\ eq.~\eqref{eq: new_general_int} with $y^2(x)$ quadratic in $x$.
We will also study these in the following, and define the map $\mathscr{I}$ to act as identity in this case.}

We now turn our focus to the pentabox integral, specifically the symbol of $\mathscr I\bigl[\widehat{\mathcal I}[\PBnoN]^{(\pm)}\bigr]$. 
Symbols of the general type $\mathcal{S}(\mathscr{I}[\tilde{\mathcal{I}}])$ are highly constrained due to their double periodicity in terms of the torus coordinates $w(c)$ and $\{w({u_a})\}$; more precisely, it follows from eq.\ \eqref{eq:kernels} that
\begin{align} \label{eq: double_periodicity_pb}
	\delta \mathcal{S}(\mathscr{I}[\tilde{\mathcal{I}}])=
	\begin{cases}
	2\mathcal{S}\Bigl(\frac{2\pi i}{\omega_{1}} \widehat{\cI}[\DB]\Bigr) & \text{for }
	\delta w(c) = \tau, \\
	0 & \text{for } \delta w(\bullet\neq c) = \tau\,.
	\end{cases}
\end{align}
Inspired by the structure of the double-box symbol, together with the integral representation \eqref{eq: int_rep_pb} and  first-two-entries and double-periodicity requirement, it is natural to guess that the $\Delta_{2,2}$-coproduct for the two pentabox integrals is
\begin{align} \label{eq: pentabox_coproduct}
	&\Delta_{2,2}\Big(\mathscr{I}\big[{-}\widehat{\mathcal I}[\PBnoN]^{(\pm)}\big] \Big)=\nonumber
	\\[-18pt]
	&\!\!\!\!\!\sum_{i<j<k}\hspace{-3pt}\widehat{\cI}[\SB]^\fourd_{[\hat{\imath}\hat{\jmath}\hat{k}]}\! \otimes \!
	\Big(\!(-1)^{k}\!\overbrace{\mathscr I\!\Big[\!\int^{u_{2}}\!\!\!\!\frac{dx~y_k(r_\pm)}{y_k(x)(x-r_\pm)} \log R_{ij}^{[\hat{k}]}(x)\Big]}^{\qquad\qquad =:T_{ij}^{[\hat{k}]}(r_\pm)}\nonumber\\[-5pt]
	&\hspace{90pt}+\text{cyc}(i,j,k)\Big)- (u_{2}\to \infty)\,. 
\end{align}
Here, the superscript $[\hat k]$ indicates the index set for the respective hexagon last entry. 
Furthermore, we use $S_{ij}^{[\hat{k}]}$ to denote the double-box counterpart of $T_{ij}^{[\hat{k}]}(r_\pm)$; cf.\ eq.~\eqref{eq: coproduct_db}.

The symbol of the new object $T_{ij}^{[\hat{k}]}(r_\pm)$ for $k\in L$ is
\begin{align} \label{eq: symbol_sijk_easy}
	&\mathcal{S}\big(T_{ij}^{[\hat{k}]}(r_\pm)\big) = \log R^{[\hat{k}]}_{ij}(u_{2})\!\otimes \Omega^{\pm}_{k}(u_{2})
	+\!  \tfrac{1}{\pi i}\partial_{\tau}S_{ij}^{[\hat{k}]}\!\otimes 2\pi i w_k(r_\pm)
	\nonumber \\
	&\!\!-\!\sum_{\substack{l=i,j\\\sigma=\pm}}(\delta_{li}+\sigma\delta_{lj})\phi(p^\sigma_{lk})\otimes \Omega^{\pm}_{k}(p_{lk}^\sigma)+\phi(r_\pm)\otimes \log R^{[\hat{k}]}_{ij}(r_\pm)\nonumber\\[-16pt]
\end{align}
for $\{i,j\}\subset L$ (or $\{i,j\}\subset R$), with $\{i,j\}$ cyclically ordered in $L$ (or $R$), and
\begin{align} \label{eq: symbol_sijk_hard}
	&\!\!\mathcal{S}(T_{ij}^{[\hat{k}]}(r_\pm)) =\log R_{ij}^{[\hat{k}]}(u_{2})\!\otimes\Omega_{k}^{\pm}(u_{2})+
	\!\tfrac{1}{\pi i}\partial_{\tau}S_{ij}^{[\hat{k}]}\!\otimes 2\pi i w_k(r_\pm) \nonumber \\
	&\!\!-\!\sum_{\substack{l=i,j\\\sigma=\pm}}\phi^{[\hat{k}]}_{ij}(p_{lk}^\sigma)\otimes
	 \Omega^{\pm}_{k}(p_{lk}^\sigma)-(-1)^{i'+j}\phi^{[\hat{k}]}_{ij}(0)\otimes\Omega^{\pm}_{k}(0)\nonumber\\[-16pt]&\hspace{85pt}+\phi^{[\hat{k}]}_{ij}(r_\pm)\otimes \log R^{[\hat{k}]}_{ij}(r_\pm)
\end{align}
for $i\in L$ and $j\in R$.
Here, we introduced $\Omega_{k}^{\pm}(x):= \tilde\Omega_k(w_k(x),w_k(r_{\pm}))$ with the subscript $k$ indicating the torus associated to the elliptic curve $y_{k}(x)$, $i'$ is the ordinal number of $i$ in the set $[\hat{k}]$, 
$p_{lk}^{\pm}$ are the two roots of $\widehat{\cG}_{[\hat{l}\hat{k}]}(x)$, 
$\phi(c)\!:=\log(1-c/u_{2})$,
and finally
\begin{equation}
\phi_{ij}^{[\hat{k}]}(c):= \int_{\beta}^{u_{2}}\!\!\! \frac{dx~ \sqrt{\widehat{\mathcal{\cG}}_{\smash{[\hat{\imath}\hat{\jmath}\hat{k}]}}(c)}}{(x-c)\sqrt{\widehat{\mathcal{\cG}}_{\smash{[\hat{\imath}\hat{\jmath}\hat{k}]}}(x)}}
\end{equation}
with $\beta$ denoting either of the two roots\footnote{The difference between the two choices integrates to $i\pi$ and hence is invisible at the symbol level.} of $\widehat{\cG}_{[\hat{\imath}\hat{\jmath}\hat{k}]}(x)$ and which integrates to a logarithm. 
The case $k\in R$ is polylogarithmic and is discussed in App.\ \ref{app: Polylog Symbol}.

Expression \eqref{eq: pentabox_coproduct}, together with eqs.\ \eqref{eq: symbol_sijk_easy} and \eqref{eq: symbol_sijk_hard}, passes several consistency checks. 
Next to satisfying relation \eqref{eq: int_rep_pb} and the first-two-entries condition by construction, also the double periodicity \eqref{eq: double_periodicity_pb} is manifest: 
for $\delta w_k(r_{\pm})= \tau_k$ one finds $\delta\Omega_{k}^{\pm}(\bullet)= 4\pi i w_{k}(\bullet)$ and thus $\delta T_{ij}^{[\hat{k}]}=2S_{ij}^{[\hat{k}]}$, while for $\delta w_k(\bullet\neq r_{\pm})=  \tau_k$ one finds $\delta\Omega_{k}^{\pm}(\bullet)=4\pi i w_k(r_{\pm})$ and thus $\delta T_{ij}^{[\hat{k}]}=0$. 
The last terms in eqs.~\eqref{eq: symbol_sijk_easy} and \eqref{eq: symbol_sijk_hard} reflect the fact that the $T_{ij}^{[\hat{k}]}$ individually become singular as $u_{2}\rightarrow r_{\pm}$, but these terms cancel in eq.\ \eqref{eq: pentabox_coproduct} due to 
\begin{equation}\label{eq:R_cancel}
	\Bigl((-1)^{k}\log R_{ij}^{[\hat{k}]}(x)+\text{cyc}(i,j,k)\Bigr)\Big\vert_{x\to u_{2},r_{\pm}}=0\,,
\end{equation}
cf.\ the discussion around eq.\ \eqref{eq:sing_cancel}. 
This relation additionally manifests the integrability condition\footnote{This condition is necessary for a symbol to be associated to a function, similar to the symmetry of second derivatives of a differential form, cf.\ refs.\ \cite{Chen:1977oja,Brown:2009qja,Duhr:2012fh}.} 
 on the middle pair of the symbol associated to eq.\ \eqref{eq: pentabox_coproduct}:
all terms but the last one 
in eqs.~\eqref{eq: symbol_sijk_easy} and \eqref{eq: symbol_sijk_hard} 
were already present in the double-box symbol, cf.\ App.\ \ref{app: symbs}, merely with a modified last entry, and thus the integrability condition here immediately follows from that of the double-box case.

Finally, let us mention that in the double-pentagon case, the only new ingredient is an integral of the form \eqref{eq: new_general_int} with $c=0$. Its $\Delta_{2,2}$-coproduct and symbol can be easily obtained from the pentabox expressions by taking $r_{\pm}\to 0$; see App.~\ref{app: Limit Symbol} for the explicit results.

\section{Discussion and Outlook}

In this letter, we have calculated the general pentabox and double-pentagon integrals, which together with the previously calculated double-box integral give all two-loop amplitudes in planar $\mathcal{N}=4$ SYM theory.
Concretely, we derived one-fold integral representations for the pentabox and double-pentagon integrals by generalising the differential-equation approach of refs.\ \cite{Paulos:2012nu, Nandan:2013ip}.
From these integral representations, we were essentially able to write down the corresponding symbols as the most naive expressions that satisfy basic consistency requirements! These expressions hold true because of significant cancellations in four-dimensional kinematics.

It remains unknown whether these integrals can be evaluated as eMPLs, but the one-fold integral representation offers a promising alternative. 
This representation provides an excellent starting point for computing the symbol and thus for studying the singularity structure of Feynman integrals,
as demonstrated in refs.\ \cite{Morales:2022csr, He:2023qld} and here. 
For numerical evaluations, it has been shown \cite{Henn:2024ngj} that one-fold integral representations also offer significant improvements
for some degenerate cases of the integrals discussed here. However, achieving a complete numerical evaluation for the most general case requires a detailed study of the integration contour as well as the analytic continuation of the one-loop hexagon into various kinematic regions, which we leave for future work.

The entire amplitude \eqref{eq: generic two loop} also includes integrals where some massive legs become massless or soft, potentially introducing IR divergences. 
The one-fold integral representation, together with the DCI regulator~\cite{Bourjaily:2013mma}, has already provided a convenient solution for the double box, as shown in refs.\ \cite{Wilhelm:2022wow,He:2022ujv}. 
The pentabox and double pentagon do not introduce additional complexity, as they only become divergent when the massive legs on the box side become massless.

The two-loop results presented here pave the way for numerous new computations in planar $\mathcal{N}=4$ SYM theory and even in QCD. 
Beyond these
 immediate applications, just as the BCF and BCFW relations~\cite{Britto:2004ap,Britto:2005fq} provided deep insights from one-loop amplitudes, 
we are more keen to ask: what new understandings about the structure of perturbative Quantum Field Theory can be gained from these two-loop results?

\begin{acknowledgments}

We thank Claude Duhr for fruitful discussions and for comments on the manuscript.
We are grateful to Zhenjie Li and Lecheng Ren for inspiring discussions on the symbol integration. AS and CZ gratefully acknowledge support from the Simons Center for Geometry and Physics, Stony Brook University, where some of the research for this paper was performed.
This work was supported by the research grant 00025445 from Villum Fonden and the ERC Starting Grant 757978. AS has furthermore received funding from the European Union's Horizon 2020 research and innovation program under the Marie Sk\l odowska-Curie grant agreement No.\ 847523 `INTERACTIONS', as well as from the Deutsche Forschungsgemeinschaft (DFG, German Research Foundation) -- Projekt\-nummer 417533893/GRK2575 ``Rethinking Quantum Field Theory''. The work of CZ is also funded by the European Union (ERC Consolidator Grant LoCoMotive 101043686). Views and opinions expressed are however those of the author(s) only and do not necessarily reflect those of the European Union or the European Research Council. Neither the European Union nor the granting authority can be held responsible for them.

\end{acknowledgments}

\appendix

\section{Integral representations} \label{app: Notations}

In this appendix, we provide further details on the one-fold integral representations given in Sec.\ \ref{sec: DEs and integral reps}.

The integral measure on the projective light cone is defined as
\begin{equation}
	\int D^{d}X := \int \frac{2 \delta (X^2) \: d^{d+2}X }{\pi^{d/2}\operatorname{Vol} \operatorname{GL}(1,\mathbb{R})^{+}}\,,
\end{equation}
such that 
\begin{equation}
\Gamma(d)\int \frac{D^{d}X}{(X,Y)^{d}} = \frac{\Gamma(d/2)}{(-Y\cdot Y)^{d/2}}\,.
\end{equation}

With this integral measure, 
 the Feynman-parameter representations for the integrals in the main text are as follows:
\begin{align}
	\cI[\DB]^{\fourd} &= \int_{[0,\infty]^{5}}\frac{[d^{2}\vec{\alpha}_{L}]\,d^{3}\vec{\alpha}_{R}}{(-Y_{L}\cdot Y_{L})(-Y_{[6]}\cdot Y_{[6]})^2}\,, \\
	\cI[\HEX]^{\sixd} &= \int_{[0,\infty]^{5}}\frac{2\,[d^{5}\vec{\alpha}_{[6]}]}{(-Y_{[6]}\cdot Y_{[6]})^{3}}\,,\\ 
	\cI[\PB]^{\fourd} &=\sum_{r\in R}\int_{[0,\infty]^{6}}\frac{2(N,X_{r})\alpha_{r}\,[d^{4}\vec{\alpha}_{L}]d^{3}\vec{\alpha}_{R}}{(-Y_{L}\cdot Y_{L})(-Y_{[7]}\cdot Y_{[7]})^3} \,,\\
	\cI[\HEP]^{\eightd} &= \int_{[0,\infty]^{6}}\frac{6\alpha_{3}\, [d^{6}\vec{\alpha}_{[7]}]}{(-Y_{[7]}\cdot Y_{[7]})^{4}} \,,\\
	\cI[\DP]^{\fourd} &= \int_{[0,\infty]^{7}} \frac{2\,[d^{3}\vec{\alpha}_{L}]d^{4}\vec{\alpha}_{R}}{(-Y_{L}\cdot Y_{L})(-Y_{[8]}\cdot Y_{[8]})^{3}} \\
	&\quad \times \left(3\frac{(N_{L},Y_{R})(N_{R},Y_{L})}{-Y_{[8]}\cdot Y_{[8]}}-(N_{L},N_{R})\right)\nonumber \,,\\
	\cI[\DPsix]^{\sixd} &= \int_{[0,\infty]^{7}}\frac{2[d^{3}\vec{\alpha}_{L}]d^{4}\vec{\alpha}_{R}}{(-Y_{L}\cdot Y_{L})(-Y_{[8]}\cdot Y_{[8]})^3} \,,\\
	\cI[\OCT]^{\eightd}&=\int_{[0,\infty]^{7}}\frac{6\, [d^{7}\vec{\alpha}_{[8]}]}{(-Y_{[8]}\cdot Y_{[8]})^{4}}\,,
\end{align}
where we have introduced $Y_{I}:=\sum_{i\in I}\alpha_{i}X_{i}$ and $d^{\lvert I\rvert}\vec{\alpha}_{I}:=\prod_{i\in I}d\alpha_{i}$
for the index set $I=L,R$ or $[n]$. Additionally, we define $[d^{|I|-1}\vec{\alpha}_{I}]:=d^{|I|}\vec{\alpha}_{I} \delta(\alpha_{i}-1)$ with arbitrary $i\in I$.

With the above Feynman-parameter representations, one can easily verify the differential equations \eqref{eq:dbDE} and \eqref{eq:pbDE}. Alternatively, one can introduce an ancillary parameter $t$ in $-Y_{[n]}\cdot Y_{[n]}$, such that
\begin{equation}
	-Y_{[n]}^{2}(t):=-tY_{L}^{2}+(Y_{L},Y_{R})-Y_{R}^{2}\,.
\end{equation}
Correspondingly, $t$-derivatives of these $t$-deformed two-loop integrals yield $t$-deformed one-loop Feynman integrals. The resulting one-fold integrals are related to those in the main text through the variable substitution $t=x/u_{2(3)}$. This approach avoids the argument about kinematic constraints and is more convenient for numerical purposes.

For a Gram determinant $\cG_{I}$ with the ordered index set $I=\{i_{1},\ldots,i_{n}\}$, we define its normalised version as 
\begin{equation}
	\widehat{\cG}_{I}:= \frac{\cG_{I}}{\prod_{a=1}^{n} (X_{i_a},X_{i_{a+\lfloor n/2\rfloor }})}\,, 
\end{equation}
where $i_{n+a}:=i_a$. We also encountered Gram determinants of the form $\cG_{[\hat{\imath}]}^{[\hat{\jmath}]}$ in the main text; their normalisations are the same as those of $\sqrt{\cG_{[n]}\cG_{[\hat{\imath}\hat{\jmath}]}}$.

The sum/difference of the two leading singularities of the pentabox in eq.\ \eqref{eq: one-fold over hexagon integrals with LS} are 
\begin{align}
	\mathsf{LS}_{\text{pb}}^{(\pm)} &= \frac{1}{2 X_{13}}\Biggl(
		\frac{(N,Q_{L}^{+})}{(X_{2},Q_{L}^{+})\sqrt{\cG_{L}\widehat{\cG}_{123Q_{L}^{+}}}} \pm (+\to -)\Biggr)\,,
\end{align}
where $Q_{L}^{\pm}$ are two null vectors orthogonal to all $X_{\ell}$ for $\ell \in L$. Moreover,
for $\{i,j\}\in R$, they satisfy 
\begin{equation}
	\frac{(Q_{L}^{\pm},X_{i})}{(Q_{L}^{\pm},X_{j})}=
	\frac{\cG_{\{4567i\}}^{\{4567j\}}\pm \langle 4567ij\rangle\sqrt{\cG_{L}}}{\cG_{\{4567j\}}}\,,
\end{equation}
where $\langle ijklmn\rangle:=\epsilon_{MNPQRS}X_{i}^{M}X_{j}^{N}X_{k}^{P}X_{l}^{Q}X_{m}^{R}X_{n}^{S}$, and 
$\langle ijklmn\rangle^{2}=-\cG_{\{ijklmn\}}$.

In eq.\ \eqref{eq: int_rep_dp}, the prefactor $\mathsf{LS}_{\text{pb}[\hat{\imath}]}^{(\pm)}$ is given for $i\in L$ by
\begin{align}
     \mathsf{LS}_{\text{pb}[\hat\imath]}^{(\pm)} &= \frac{X_{57}X_{68}}{2}\Biggl(
     \frac{(N_{L},Q_{R}^{+})(N_{R},Q_{R}^{+})\widehat{\cG}_{L}^{L_{i}^{+}}}{\cG_{L^{+}}\sqrt{\cG_R
 	\widehat{\cG}_{L_{i}^{+}}}} \pm (+\to -)\!\Biggr) \,,
\end{align}
where the normalisation of $\cG_{L}^{L_{i}^{+}}$ is the same as $\sqrt{\cG_{L}\cG_{L_{i}^{+}}}$, and the index sets are $L^{\pm}:=L\cup\{Q_{R}^{\pm}\}$ and $L_{i}^{\pm}=L^{\pm}\setminus \{i\}$. The result for $i\in R$ can be easily obtained by the reflection $i\to 9-i$. 
For the kissing-box configuration, the prefactor is given by 
\begin{equation}
	\mathsf{LS}_{\text{kb}}=\frac{1}{4}\!\!\sum_{(\sigma_{1},\sigma_{2})\in (\pm,\pm)}\!\!\!\!
	\sigma_{1}\sigma_{2}\frac{(N_{L},Q_{L}^{\sigma_{1}})(N_{R},Q_{R}^{\sigma_{2}})}{\sqrt{\cG_{L}\cG_{R}}(Q_{L}^{\sigma_{1}},Q_{R}^{\sigma_{2}})}\,.
\end{equation}
Moreover, our convention for $y_{\ell r}(0)$ is 
\begin{equation}
	y_{\ell r}(0)
	=\frac{\cG_{\{\ell_{1}\ell_{2}\ell_{3}\}}^{\{r_{1}r_{2}r_{3}\}}}{N_{\ell r}} \quad \text{with} \quad N_{\ell r}= \prod_{i=1}^{3} X_{\ell_{i},r_{i}}\,,
\end{equation}
where $\{\ell_{1}\ell_{2}\ell_{3}\}=L\setminus\{\ell\}$ and likewise for $\{r_{1}r_{2}r_{3}\}$.
To compute $C_{ij}$, we introduce two quadratic polynomials in $x$, 
\begin{align}
	P_{2}(x)&:=(x-u_{3})^{-2}\widehat{\cG}_{[8]}(x)\,,\\
	Q_{2}(x)&:=\sum_{\ell\in L,r\in R}\frac{(-1)^{\ell+r}(N_{L},X_{r})(N_{R},X_{\ell})}{(N_{L},N_{R})\prod_{i=1}^{4}X_{i,i+4}}
	N_{\ell r} \frac{\widehat{\cG}_{[\hat{r}]}^{[\hat{\ell}]}(x)}{x-u_{3}}\nonumber
\end{align}
and their $x$-derivatives $P'_{2}(x)$, $P_{2}''(x)$, etc.
Then, 
\begin{align}
	C_{\ell r}&= \frac{k (N_{L},N_{R})}{X_{14}X_{26}X_{37}X_{58}}\lim_{x\to\infty} \frac{(x-u_{3})\widehat{\cG}_{[\hat{r}]}^{[\hat{\ell}]}(x)}{2\widehat{\cG}_{[8]}(x)}\,,
\end{align}
where
\begin{equation}
	k= \frac{P_{2}''(x)\bigl(P_{2}'(x)-Q_{2}''(x)\bigr)}{\bigl(P_{2}'(x)\bigr)^{2}-2P_{2}(x)P_{2}''(x)}\Bigg\vert_{x\to u_{3}}\,.
\end{equation}

\section{Symbol building blocks}
\label{app: symbs}

In this appendix, we collect the (known) symbols for the one-loop box and double box, which are building blocks for the 
pentabox and double-pentagon symbols.

The result for the one-loop box integral with the particle set $\{i,j,k,l\}$ is 
\begin{align}
	\widehat{\cI}\big[\SB\big]^{\fourd}_{\{ijkl\}} &:= \tfrac{1}{2}\sqrt{\cG_{\{ijkl\}}}\cI\big[\SB\big]^{\fourd}_{\{ijkl\}} \\
	&=\operatorname{Li}_{2}(z)-\operatorname{Li}_{2}(\bar{z})+\tfrac{1}{2}\log(z\bar{z})\log\frac{1-z}{1-\bar{z}}\,,\nonumber
\end{align} 
where 
\begin{gather}
z\bar{z}=\frac{X_{ij}X_{kl}}{X_{ik}X_{jl}},\quad (1-z)(1-\bar{z})=\frac{X_{jk}X_{il}}{X_{ik}X_{jl}}\,,
\end{gather}
with $z-\bar{z}=\sqrt{\widehat{\cG}_{\{ijkl\}}}$.

Now, let us describe the symbol of $S_{ij}$ appearing in the double-box $\Delta_{2,2}$-coproduct, cf.\ eq.\ \eqref{eq: coproduct_db}.
For $\{i,j\}\subset L$ (or $\{i,j\}\subset R$), with $\{i,j\}$ cyclically ordered in $L$ (or $R$), 
\begin{align}\label{eq: Sij equal}
\mathcal S( S_{ij})=&\log R_{ij}(u_2)\otimes 2\pi i\,w(u_2)+\tfrac{1}{2\pi i}\partial_\tau S_{ij}\otimes 2\pi i\tau\\
 &\hspace{40pt}-\sum_{\substack{l=i,j\\\sigma=\pm}}(\delta_{li}+\sigma\delta_{lj})\phi(p^\sigma_l)\otimes 2\pi i\,w(p_l^\sigma)\,.\nonumber
\end{align}
If instead $i$ and $j$ take one value from each set,
e.g.\ $i=\ell \in L$ and $j=r\in R$,
then
\begin{align}
&\mathcal S(S_{\ell r})=\,\log R_{\ell r}(u_2)\otimes 2\pi i\,w(u_2)+\tfrac{1}{2\pi i}\partial_\tau S_{\ell r}\otimes 2\pi i\tau
\label{eq: Sij unequal}\\
&-\sum_{\substack{a=\ell,r\\\sigma=\pm}}\phi_{\ell r}(p_a^\sigma)\otimes 2\pi i\, w(p_a^\sigma)-(-1)^{\ell+r}\phi_{\ell r}(0)\otimes 2\pi i\,w(0)\,\nonumber
\end{align}
with 
\begin{align}
\phi_{\ell r}(c)=\int^{u_2}_{\beta}\!\!\!\frac{dx~\sqrt{\widehat\cG_{[\hat\ell \hat r]}(c)}}{(x-c)\sqrt{\widehat\cG_{[\hat\ell \hat r]}(x)}}\,.
\end{align}

For $p_{i}^{\pm}$ and their corresponding values $y(p_{i}^{\pm})$ on the elliptic curve $y^{2}=-\widehat{\cG}_{[6]}(x)$,  
we follow the convention used in ref.~\cite{Morales:2022csr}:
\begin{align}
	p_{i}^{\pm}&=u_{2}\Biggl(1-\frac{\left((-1)^{k+j}\cG_{[\hat{\imath}\hat{\jmath}]}^{_{[\hat{\imath}\hat{k}]}}\pm
	\sqrt{\cG_{[\hat{\imath}\hat{\jmath}]}\cG_{[\hat{\imath}\hat{k}]}}\right)X_{ij}X_{ik}}{2X_{12}X_{23}X_{31}X_{45}X_{56}X_{46}} \Biggr) \,, \nonumber\\
	y(p_{i}^{\pm})&=(-1)^{i+j}\widehat{\cG}_{[\hat{\imath}]}^{[\hat{\jmath}]}(x\to p_{i}^{\pm})\Big /\sqrt{\widehat{\cG}_{[\hat{\imath}\hat{\jmath}]}} \,,
\end{align}
where $\{i,j,k\}$ is one of three cyclic permutations of either $\{1,2,3\}$ or $\{4,5,6\}$, depending on the value of $i$.
As for $\phi_{\ell r}(p_{a}^{\sigma})$ in eq.~\eqref{eq: Sij unequal}, we also need to determine $\sqrt{\widehat{\cG}_{[\hat{\ell}\hat{r}]}(p_{a}^{\sigma})}$. The convention we adopted is such that
\begin{align}
	\phi_{\ell r}(p_{\ell}^{+}) \pm \phi_{\ell r}(p_{\ell}^{-})
	=\pm(-1)^{\ell+r}\log\frac{\cG^{[\hat{\ell}\hat{r}]}_{[\hat{\ell}\,\widehat{\ell\pm1}]}-\sqrt{\cG_{[\hat{\ell}\hat{r}]}\cG_{[\hat{\ell}\,\widehat{\ell\pm1}]}}}{\cG^{[\hat{\ell}\hat{r}]}_{[\hat{\ell}\,\widehat{\ell\pm1}]}+\sqrt{\cG_{[\hat{\ell}\hat{r}]}\cG_{[\hat{\ell}\,\widehat{\ell\pm1}]}}}\,, \nonumber\\
	\phi_{\ell r}(p_{r}^{+}) \pm \phi_{ij}(p_{r}^{-})
	=\mp(-1)^{\ell+r}\log\frac{\cG^{[\hat{\ell}\hat{r}]}_{[\hat{r}\,\widehat{r\pm1}]}-\sqrt{\cG_{[\hat{\ell}\hat{r}]}\cG_{[\hat{r}\,\widehat{r\pm1}]}}}{\cG^{[\hat{\ell}\hat{r}]}_{[\hat{r}\,\widehat{r\pm1}]}+\sqrt{\cG_{[\hat{\ell}\hat{r}]}\cG_{[\hat{r}\,\widehat{r\pm1}]}}}\,.
\end{align}
Here, the indices $\ell\pm 1$ should be understood as the cyclic successor/predecessor of $\ell$ in the set $L$, 
and likewise for $r$.

Note that the last entries $w(p_{l}^{\sigma})$ are not independent; however, the integrability condition for the symbol of the double box is satisfied regardless of this fact.

\section{Polylogarithmic symbols}

\label{app: Polylog Symbol}  

In this appendix, we present the symbols associated to 
the integrals $T_{ij}^{[\hat k]}(r_\pm)$ defined in the coproduct of the pentabox integrals \eqref{eq: pentabox_coproduct} for $k\in R$, which are polylogarithmic. The results are actually independent of the particular values of $r_{\pm}$; thus, we will use $T_{ij}^{[\hat k]}(c)$ with an arbitrary $c$ below to simplify the notation.

\textbf{Case (i)}: $\{i,j\}=R\backslash\{k\}$. 
This case has no additional square roots arising from the four-mass box associated to indices $[\hat\imath\hat\jmath\hat k]$, 
and one can find 
\begin{align}
\mathcal{S}\big(T_{ij}^{[\hat k]}(c)\big)=&~\log R_{ij}^{[\hat{k}]}(u_2) \otimes \Phi_{k}(u_{2})\\
	&\qquad -2 \log\frac{4 (c-u_{2})}{r^{[\hat{k}]}_{+}-r^{[\hat{k}]}_{-}} 
	\otimes \Phi_{k}(\infty)\,, \nonumber
\end{align}
where 
\begin{align}
	\Phi_{k}(a)&=\! \int^{c}_{r^{[\hat{k}]}_{\pm}}\! \frac{d x~y_k(a)}{(x-a)y_k(x)} 
	=\!\! \int^{a}_{r^{[\hat{k}]}_{\pm}}\!\! \frac{dx~y_{k}(c)}{(x-c)y_{k}(x)}\,,\\
	\Phi_{k}(\infty)&=	-\frac{1}{2}\log R_{ij}^{[\hat{k}]}(c)
\end{align}
and $y_k(x)$ is quadratic in $x$, with roots in four-dimensional kinematics given by
\begin{equation}
	r^{[\hat{k}]}_{\pm} = u_{2}\left(
	1+\frac{-\mathcal G_{[\hat{k}\hat{\imath}]}^{[\hat{k}\hat{\jmath}]}\pm \sqrt{\mathcal G_{[\hat{k}\hat{\imath}]}\mathcal G_{[\hat{k}\hat{\jmath}]}}}{\mathcal G_{[\hat{k}\hat{\imath}\hat{\jmath}]}X_{ij}}
	\right)\,.
\end{equation}

\textbf{Case (ii)}: $i \in R\backslash \{k\}$ and $j\in L$. 
This case has no additional square roots as well, and one can find\allowdisplaybreaks[0]
\begin{align}
&\mathcal{S}\big(T_{ij}^{[\hat k]}(c) \big)= \log R_{ij}^{[\hat k]}(u_2) \otimes \Phi_{k}(u_{2}) +\phi(c) \otimes \log R^{[\hat{k}]}_{ij}(c)\nonumber \\
	&\qquad - \sum_{\sigma=\pm} (\delta_{i,k+1}+\sigma\delta_{i,k-1})~\phi(p_{kj}^{\sigma}) \otimes \Phi_{k}(p_{kj}^{\sigma})\nonumber\\[-0.25\baselineskip]
	&\qquad -
	(-1)^{j}\log \frac{\cG_{[\hat{\imath}\hat{\jmath}\hat{k}]}^{[\hat{1}\hat{2}\hat{3}]}
	-\sqrt{\cG_{[\hat{\imath}\hat{\jmath}\hat{k}]}\cG_{[\hat{1}\hat{2}\hat{3}]}}}{\cG_{[\hat{\imath}\hat{\jmath}\hat{k}]}^{[\hat{1}\hat{2}\hat{3}]}+\sqrt{\cG_{[\hat{\imath}\hat{\jmath}\hat{k}]}\cG_{[\hat{1}\hat{2}\hat{3}]}}} \otimes \Phi_{k}(\infty) \,, 
\end{align}
Similar to the case of the double box, we need to determine $y_{k}(p_{kj}^{\sigma})$ to compute $\Phi_{k}(p_{kj}^{\sigma})$, and the convention we adopted is such that
\begin{equation}
	\Phi_{k}(p_{kj}^{+})\pm 	\Phi_{k}(p_{kj}^{-})
	=\log R^{[\hat{k}]}_{k\pm1,j}(c)\,,
\end{equation}
where $k\pm 1$ should again be understood as the cyclic successor/predecessor of $k$ in the set $L$.

\textbf{Case (iii)}: $\{i,j\} \subset L$. 
This case involves an additional square root arising from the associated four-mass box, and we find
\begin{align}
&\mathcal{S}\big(T^{[\hat{k}]}_{ij}(c) \big)
=\log R_{ij}^{[\hat{k}]}(u_2) \otimes \Phi_{k}(u_{2}) \nonumber \\
	&\quad -\sum_{\substack{l=i,j\\\sigma=\pm}}
	\phi_{ij}^{[\hat{k}]}(p_{kl}^{\sigma}) \otimes 
	\Phi_{k}(p_{lk}^{\sigma}) + \phi_{ij}^{[\hat{k}]}(c) \otimes \log R_{ij}^{[\hat{k}]}(c)\,.
\end{align}
Note that $\phi_{ij}^{[\hat{k}]}(p_{kl}^{\sigma})$ are not new objects; they can be computed by, e.g., $\phi_{ij}^{[\hat{k}]}(p_{ki}^{\sigma})=\phi_{kj}^{[\hat{\imath}]}(p_{ki}^{\sigma})$, which we have already discussed in the case of the double box in App.~\ref{app: symbs}.

\vspace{\baselineskip}

\section{Limits of symbols}
\label{app: Limit Symbol}  

In this final appendix, let us describe the symbols $\mathcal{S}(T_{ij}^{[\hat{\ell} \hat{r}]}(0))$, 
 which are the new ingredients appearing in the double pentagon.
More precisely, we have
\begin{align}
	&\Delta_{2,2}\left(\mathscr{I}\biggl[-\int_{u_{3}}^{\infty}\frac{dx~y_{\ell r}(0)}{x~y_{\ell r}(x)}\widehat{\cI}[\HEX]^{\sixd}_{[\hat{\ell}\hat{r}]}\biggr]\right)  \\[-0.75\baselineskip]
	&= \sum_{\{i,j\}\subset [\hat{\ell}\hat{r}]} \widehat{\cI}[\SB]^{\fourd}_{[\hat{\imath}\hat{\jmath}\hat{\ell}\hat{r}]} \otimes \overbrace{\mathscr I\!\Big[\!\int^{u_{3}}\!\frac{dx~y_{\ell r}(0)}{x\,y_{\ell r}(x)} \log R_{ij}^{[\hat{\ell}\hat{r}]}(x)\Big]}^{\qquad\qquad =:T_{ij}^{[\hat{\ell}\hat{r}]}(0)}\,.\nonumber
\end{align}
There are two cases: (i) $\{i,j\}$ is a subset of $L$ or $R$, and (ii) $i\in L$ and $j\in R$. The case (i) is trivial; one can simply obtain the result by taking $r_{\pm}=0$ in eq.~\eqref{eq: symbol_sijk_easy}. Note that, the last term in eq.~\eqref{eq: symbol_sijk_easy} vanishes due to $\phi(0)=0$ and then the integrability condition holds as we argued at the end of Sec.~\ref{sec:dbSym}. The case (ii) is less trivial; the last two terms in eq.~\eqref{eq: symbol_sijk_hard} become divergent as $r_{\pm}\to 0$, but these divergences cancel each other. As a result, we have
\begin{align}
	&\lim_{r_{\pm}\to 0} -\Omega_{\ell r}^{\pm}(0)+(-1)^{i'+j'}\log R_{ij}^{[\hat{\ell}\hat{r}]}(r_{\pm})
	\nonumber \\ 
	&=4\log\frac{\theta_{1}(w_{\ell r}(0))}{\eta(\tau_{\ell r})} +2\log\frac{a_{\ell r}}{4}-\tfrac{1}{3}\log \Delta_{\ell r} + \cdots\,, 
\end{align}
where $i'$ and $j'$ are the ordinal numbers of $i$ and $j$ in the set $[\hat{\ell}\hat{r}]$, 
$a_{\ell r}$ is the leading coefficient of $-\widehat{\cG}_{[\hat{\ell}\hat{r}]}(x)$ in $x$, 
$\Delta_{\ell r}=g_{2}^{3}-27 g_{3}^{2}$ is the discriminant obtained from the standard Weierstrass form $Y^{2}=4X^{3}-g_{2}X-g_{3}$ of the elliptic curve 
$y^{2}_{\ell r}=-\widehat{\cG}_{[\hat{\ell}\hat{r}]}(x)$, $\eta$ is the Dedekind eta function and its argument $\tau_{\ell r}$ should be understood as the modular parameter of the elliptic curve $y^{2}_{\ell r}=-\widehat{\cG}_{[\hat{\ell}\hat{r}]}(x)$. There are also some terms which cancel out in the sum in eq.~\eqref{eq: int_rep_pb}; we omit these terms as indicated by the ellipsis.

\bibliography{reference} 

%apsrev4-2.bst 2019-01-14 (MD) hand-edited version of apsrev4-1.bst
%Control: key (0)
%Control: author (8) initials jnrlst
%Control: editor formatted (1) identically to author
%Control: production of article title (0) allowed
%Control: page (0) single
%Control: year (1) truncated
%Control: production of eprint (0) enabled
\providecommand{\noopsort}[1]{}\providecommand{\singleletter}[1]{#1}%
\begin{thebibliography}{53}%
\makeatletter
\providecommand \@ifxundefined [1]{%
 \@ifx{#1\undefined}
}%
\providecommand \@ifnum [1]{%
 \ifnum #1\expandafter \@firstoftwo
 \else \expandafter \@secondoftwo
 \fi
}%
\providecommand \@ifx [1]{%
 \ifx #1\expandafter \@firstoftwo
 \else \expandafter \@secondoftwo
 \fi
}%
\providecommand \natexlab [1]{#1}%
\providecommand \enquote  [1]{``#1''}%
\providecommand \bibnamefont  [1]{#1}%
\providecommand \bibfnamefont [1]{#1}%
\providecommand \citenamefont [1]{#1}%
\providecommand \href@noop [0]{\@secondoftwo}%
\providecommand \href [0]{\begingroup \@sanitize@url \@href}%
\providecommand \@href[1]{\@@startlink{#1}\@@href}%
\providecommand \@@href[1]{\endgroup#1\@@endlink}%
\providecommand \@sanitize@url [0]{\catcode `\\12\catcode `\$12\catcode `\&12\catcode `\#12\catcode `\^12\catcode `\_12\catcode `\%12\relax}%
\providecommand \@@startlink[1]{}%
\providecommand \@@endlink[0]{}%
\providecommand \url  [0]{\begingroup\@sanitize@url \@url }%
\providecommand \@url [1]{\endgroup\@href {#1}{\urlprefix }}%
\providecommand \urlprefix  [0]{URL }%
\providecommand \Eprint [0]{\href }%
\providecommand \doibase [0]{https://doi.org/}%
\providecommand \selectlanguage [0]{\@gobble}%
\providecommand \bibinfo  [0]{\@secondoftwo}%
\providecommand \bibfield  [0]{\@secondoftwo}%
\providecommand \translation [1]{[#1]}%
\providecommand \BibitemOpen [0]{}%
\providecommand \bibitemStop [0]{}%
\providecommand \bibitemNoStop [0]{.\EOS\space}%
\providecommand \EOS [0]{\spacefactor3000\relax}%
\providecommand \BibitemShut  [1]{\csname bibitem#1\endcsname}%
\let\auto@bib@innerbib\@empty
%</preamble>
\bibitem [{\citenamefont {Bern}\ \emph {et~al.}(1994)\citenamefont {Bern}, \citenamefont {Dixon}, \citenamefont {Dunbar},\ and\ \citenamefont {Kosower}}]{Bern:1994zx}%
  \BibitemOpen
  \bibfield  {author} {\bibinfo {author} {\bibfnamefont {Z.}~\bibnamefont {Bern}}, \bibinfo {author} {\bibfnamefont {L.~J.}\ \bibnamefont {Dixon}}, \bibinfo {author} {\bibfnamefont {D.~C.}\ \bibnamefont {Dunbar}},\ and\ \bibinfo {author} {\bibfnamefont {D.~A.}\ \bibnamefont {Kosower}},\ }\bibfield  {title} {\bibinfo {title} {{One loop $n$ point gauge theory amplitudes, unitarity and collinear limits}},\ }\href {https://doi.org/10.1016/0550-3213(94)90179-1} {\bibfield  {journal} {\bibinfo  {journal} {Nucl. Phys. B}\ }\textbf {\bibinfo {volume} {425}},\ \bibinfo {pages} {217} (\bibinfo {year} {1994})},\ \Eprint {https://arxiv.org/abs/hep-ph/9403226} {arXiv:hep-ph/9403226} \BibitemShut {NoStop}%
\bibitem [{\citenamefont {Bern}\ \emph {et~al.}(1995)\citenamefont {Bern}, \citenamefont {Dixon}, \citenamefont {Dunbar},\ and\ \citenamefont {Kosower}}]{Bern:1994cg}%
  \BibitemOpen
  \bibfield  {author} {\bibinfo {author} {\bibfnamefont {Z.}~\bibnamefont {Bern}}, \bibinfo {author} {\bibfnamefont {L.~J.}\ \bibnamefont {Dixon}}, \bibinfo {author} {\bibfnamefont {D.~C.}\ \bibnamefont {Dunbar}},\ and\ \bibinfo {author} {\bibfnamefont {D.~A.}\ \bibnamefont {Kosower}},\ }\bibfield  {title} {\bibinfo {title} {{Fusing gauge theory tree amplitudes into loop amplitudes}},\ }\href {https://doi.org/10.1016/0550-3213(94)00488-Z} {\bibfield  {journal} {\bibinfo  {journal} {Nucl. Phys. B}\ }\textbf {\bibinfo {volume} {435}},\ \bibinfo {pages} {59} (\bibinfo {year} {1995})},\ \Eprint {https://arxiv.org/abs/hep-ph/9409265} {arXiv:hep-ph/9409265} \BibitemShut {NoStop}%
\bibitem [{\citenamefont {Britto}\ \emph {et~al.}(2005{\natexlab{a}})\citenamefont {Britto}, \citenamefont {Cachazo},\ and\ \citenamefont {Feng}}]{Britto:2004ap}%
  \BibitemOpen
  \bibfield  {author} {\bibinfo {author} {\bibfnamefont {R.}~\bibnamefont {Britto}}, \bibinfo {author} {\bibfnamefont {F.}~\bibnamefont {Cachazo}},\ and\ \bibinfo {author} {\bibfnamefont {B.}~\bibnamefont {Feng}},\ }\bibfield  {title} {\bibinfo {title} {{New recursion relations for tree amplitudes of gluons}},\ }\href {https://doi.org/10.1016/j.nuclphysb.2005.02.030} {\bibfield  {journal} {\bibinfo  {journal} {Nucl. Phys. B}\ }\textbf {\bibinfo {volume} {715}},\ \bibinfo {pages} {499} (\bibinfo {year} {2005}{\natexlab{a}})},\ \Eprint {https://arxiv.org/abs/hep-th/0412308} {arXiv:hep-th/0412308} \BibitemShut {NoStop}%
\bibitem [{\citenamefont {Britto}\ \emph {et~al.}(2005{\natexlab{b}})\citenamefont {Britto}, \citenamefont {Cachazo}, \citenamefont {Feng},\ and\ \citenamefont {Witten}}]{Britto:2005fq}%
  \BibitemOpen
  \bibfield  {author} {\bibinfo {author} {\bibfnamefont {R.}~\bibnamefont {Britto}}, \bibinfo {author} {\bibfnamefont {F.}~\bibnamefont {Cachazo}}, \bibinfo {author} {\bibfnamefont {B.}~\bibnamefont {Feng}},\ and\ \bibinfo {author} {\bibfnamefont {E.}~\bibnamefont {Witten}},\ }\bibfield  {title} {\bibinfo {title} {{Direct proof of tree-level recursion relation in Yang-Mills theory}},\ }\href {https://doi.org/10.1103/PhysRevLett.94.181602} {\bibfield  {journal} {\bibinfo  {journal} {Phys. Rev. Lett.}\ }\textbf {\bibinfo {volume} {94}},\ \bibinfo {pages} {181602} (\bibinfo {year} {2005}{\natexlab{b}})},\ \Eprint {https://arxiv.org/abs/hep-th/0501052} {arXiv:hep-th/0501052} \BibitemShut {NoStop}%
\bibitem [{\citenamefont {Arkani-Hamed}\ \emph {et~al.}(2016)\citenamefont {Arkani-Hamed}, \citenamefont {Bourjaily}, \citenamefont {Cachazo}, \citenamefont {Goncharov}, \citenamefont {Postnikov},\ and\ \citenamefont {Trnka}}]{ArkaniHamed:2012nw}%
  \BibitemOpen
  \bibfield  {author} {\bibinfo {author} {\bibfnamefont {N.}~\bibnamefont {Arkani-Hamed}}, \bibinfo {author} {\bibfnamefont {J.~L.}\ \bibnamefont {Bourjaily}}, \bibinfo {author} {\bibfnamefont {F.}~\bibnamefont {Cachazo}}, \bibinfo {author} {\bibfnamefont {A.~B.}\ \bibnamefont {Goncharov}}, \bibinfo {author} {\bibfnamefont {A.}~\bibnamefont {Postnikov}},\ and\ \bibinfo {author} {\bibfnamefont {J.}~\bibnamefont {Trnka}},\ }\href {https://inspirehep.net/record/1208741/files/arXiv:1212.5605.pdf} {\emph {\bibinfo {title} {{Grassmannian Geometry of Scattering Amplitudes}}}}\ (\bibinfo  {publisher} {Cambridge University Press},\ \bibinfo {year} {2016})\ \Eprint {https://arxiv.org/abs/1212.5605} {arXiv:1212.5605 [hep-th]} \BibitemShut {NoStop}%
%%CITATION = ARXIV:1212.5605;%%
\bibitem [{\citenamefont {Henn}\ and\ \citenamefont {Plefka}(2014)}]{Henn:2014yza}%
  \BibitemOpen
  \bibfield  {author} {\bibinfo {author} {\bibfnamefont {J.~M.}\ \bibnamefont {Henn}}\ and\ \bibinfo {author} {\bibfnamefont {J.~C.}\ \bibnamefont {Plefka}},\ }\href {https://doi.org/10.1007/978-3-642-54022-6} {\emph {\bibinfo {title} {{Scattering Amplitudes in Gauge Theories}}}},\ Vol.\ \bibinfo {volume} {883}\ (\bibinfo  {publisher} {Springer},\ \bibinfo {address} {Berlin},\ \bibinfo {year} {2014})\BibitemShut {NoStop}%
\bibitem [{\citenamefont {Elvang}\ and\ \citenamefont {Huang}(2015)}]{Elvang:2015rqa}%
  \BibitemOpen
  \bibfield  {author} {\bibinfo {author} {\bibfnamefont {H.}~\bibnamefont {Elvang}}\ and\ \bibinfo {author} {\bibfnamefont {Y.-t.}\ \bibnamefont {Huang}},\ }\href {https://doi.org/10.1017/CBO9781107706620} {\emph {\bibinfo {title} {{Scattering Amplitudes in Gauge Theory and Gravity}}}}\ (\bibinfo  {publisher} {Cambridge University Press},\ \bibinfo {year} {2015})\BibitemShut {NoStop}%
\bibitem [{\citenamefont {Travaglini}\ \emph {et~al.}(2022)\citenamefont {Travaglini} \emph {et~al.}}]{Travaglini:2022uwo}%
  \BibitemOpen
  \bibfield  {author} {\bibinfo {author} {\bibfnamefont {G.}~\bibnamefont {Travaglini}} \emph {et~al.},\ }\bibfield  {title} {\bibinfo {title} {{The SAGEX review on scattering amplitudes}},\ }\href {https://doi.org/10.1088/1751-8121/ac8380} {\bibfield  {journal} {\bibinfo  {journal} {J. Phys. A}\ }\textbf {\bibinfo {volume} {55}},\ \bibinfo {pages} {443001} (\bibinfo {year} {2022})},\ \Eprint {https://arxiv.org/abs/2203.13011} {arXiv:2203.13011 [hep-th]} \BibitemShut {NoStop}%
\bibitem [{\citenamefont {Bourjaily}\ \emph {et~al.}(2022)\citenamefont {Bourjaily} \emph {et~al.}}]{Bourjaily:2022bwx}%
  \BibitemOpen
  \bibfield  {author} {\bibinfo {author} {\bibfnamefont {J.~L.}\ \bibnamefont {Bourjaily}} \emph {et~al.},\ }\bibfield  {title} {\bibinfo {title} {{Functions Beyond Multiple Polylogarithms for Precision Collider Physics}},\ }in\ \href@noop {} {\emph {\bibinfo {booktitle} {{Snowmass 2021}}}}\ (\bibinfo {year} {2022})\ \Eprint {https://arxiv.org/abs/2203.07088} {arXiv:2203.07088 [hep-ph]} \BibitemShut {NoStop}%
\bibitem [{\citenamefont {Laporta}\ and\ \citenamefont {Remiddi}(2005)}]{Laporta:2004rb}%
  \BibitemOpen
  \bibfield  {author} {\bibinfo {author} {\bibfnamefont {S.}~\bibnamefont {Laporta}}\ and\ \bibinfo {author} {\bibfnamefont {E.}~\bibnamefont {Remiddi}},\ }\bibfield  {title} {\bibinfo {title} {{Analytic treatment of the two loop equal mass sunrise graph}},\ }\href {https://doi.org/10.1016/j.nuclphysb.2004.10.044} {\bibfield  {journal} {\bibinfo  {journal} {Nucl. Phys. B}\ }\textbf {\bibinfo {volume} {704}},\ \bibinfo {pages} {349} (\bibinfo {year} {2005})},\ \Eprint {https://arxiv.org/abs/hep-ph/0406160} {arXiv:hep-ph/0406160} \BibitemShut {NoStop}%
\bibitem [{\citenamefont {Adams}\ \emph {et~al.}(2013)\citenamefont {Adams}, \citenamefont {Bogner},\ and\ \citenamefont {Weinzierl}}]{Adams:2013nia}%
  \BibitemOpen
  \bibfield  {author} {\bibinfo {author} {\bibfnamefont {L.}~\bibnamefont {Adams}}, \bibinfo {author} {\bibfnamefont {C.}~\bibnamefont {Bogner}},\ and\ \bibinfo {author} {\bibfnamefont {S.}~\bibnamefont {Weinzierl}},\ }\bibfield  {title} {\bibinfo {title} {{The two-loop sunrise graph with arbitrary masses}},\ }\href {https://doi.org/10.1063/1.4804996} {\bibfield  {journal} {\bibinfo  {journal} {J. Math. Phys.}\ }\textbf {\bibinfo {volume} {54}},\ \bibinfo {pages} {052303} (\bibinfo {year} {2013})},\ \Eprint {https://arxiv.org/abs/1302.7004} {arXiv:1302.7004 [hep-ph]} \BibitemShut {NoStop}%
\bibitem [{\citenamefont {Bloch}\ and\ \citenamefont {Vanhove}(2015)}]{Bloch:2013tra}%
  \BibitemOpen
  \bibfield  {author} {\bibinfo {author} {\bibfnamefont {S.}~\bibnamefont {Bloch}}\ and\ \bibinfo {author} {\bibfnamefont {P.}~\bibnamefont {Vanhove}},\ }\bibfield  {title} {\bibinfo {title} {{The elliptic dilogarithm for the sunset graph}},\ }\href {https://doi.org/10.1016/j.jnt.2014.09.032} {\bibfield  {journal} {\bibinfo  {journal} {J. Number Theor.}\ }\textbf {\bibinfo {volume} {148}},\ \bibinfo {pages} {328} (\bibinfo {year} {2015})},\ \Eprint {https://arxiv.org/abs/1309.5865} {arXiv:1309.5865 [hep-th]} \BibitemShut {NoStop}%
\bibitem [{\citenamefont {Broedel}\ \emph {et~al.}(2018{\natexlab{a}})\citenamefont {Broedel}, \citenamefont {Duhr}, \citenamefont {Dulat},\ and\ \citenamefont {Tancredi}}]{Broedel:2017siw}%
  \BibitemOpen
  \bibfield  {author} {\bibinfo {author} {\bibfnamefont {J.}~\bibnamefont {Broedel}}, \bibinfo {author} {\bibfnamefont {C.}~\bibnamefont {Duhr}}, \bibinfo {author} {\bibfnamefont {F.}~\bibnamefont {Dulat}},\ and\ \bibinfo {author} {\bibfnamefont {L.}~\bibnamefont {Tancredi}},\ }\bibfield  {title} {\bibinfo {title} {{Elliptic polylogarithms and iterated integrals on elliptic curves II: an application to the sunrise integral}},\ }\href {https://doi.org/10.1103/PhysRevD.97.116009} {\bibfield  {journal} {\bibinfo  {journal} {Phys. Rev. D}\ }\textbf {\bibinfo {volume} {97}},\ \bibinfo {pages} {116009} (\bibinfo {year} {2018}{\natexlab{a}})},\ \Eprint {https://arxiv.org/abs/1712.07095} {arXiv:1712.07095 [hep-ph]} \BibitemShut {NoStop}%
\bibitem [{\citenamefont {Bogner}\ \emph {et~al.}(2020)\citenamefont {Bogner}, \citenamefont {M\"uller-Stach},\ and\ \citenamefont {Weinzierl}}]{Bogner:2019lfa}%
  \BibitemOpen
  \bibfield  {author} {\bibinfo {author} {\bibfnamefont {C.}~\bibnamefont {Bogner}}, \bibinfo {author} {\bibfnamefont {S.}~\bibnamefont {M\"uller-Stach}},\ and\ \bibinfo {author} {\bibfnamefont {S.}~\bibnamefont {Weinzierl}},\ }\bibfield  {title} {\bibinfo {title} {{The unequal mass sunrise integral expressed through iterated integrals on $\overline{\mathcal M}_{1,3}$}},\ }\href {https://doi.org/10.1016/j.nuclphysb.2020.114991} {\bibfield  {journal} {\bibinfo  {journal} {Nucl. Phys. B}\ }\textbf {\bibinfo {volume} {954}},\ \bibinfo {pages} {114991} (\bibinfo {year} {2020})},\ \Eprint {https://arxiv.org/abs/1907.01251} {arXiv:1907.01251 [hep-th]} \BibitemShut {NoStop}%
\bibitem [{\citenamefont {Brown}\ and\ \citenamefont {Levin}(2011)}]{brown2011multiple}%
  \BibitemOpen
  \bibfield  {author} {\bibinfo {author} {\bibfnamefont {F.}~\bibnamefont {Brown}}\ and\ \bibinfo {author} {\bibfnamefont {A.}~\bibnamefont {Levin}},\ }\bibfield  {title} {\bibinfo {title} {{Multiple Elliptic Polylogarithms}},\ }\href@noop {} {\  (\bibinfo {year} {2011})},\ \Eprint {https://arxiv.org/abs/1110.6917} {arXiv:1110.6917} \BibitemShut {NoStop}%
\bibitem [{\citenamefont {Broedel}\ \emph {et~al.}(2018{\natexlab{b}})\citenamefont {Broedel}, \citenamefont {Duhr}, \citenamefont {Dulat},\ and\ \citenamefont {Tancredi}}]{Broedel:2017kkb}%
  \BibitemOpen
  \bibfield  {author} {\bibinfo {author} {\bibfnamefont {J.}~\bibnamefont {Broedel}}, \bibinfo {author} {\bibfnamefont {C.}~\bibnamefont {Duhr}}, \bibinfo {author} {\bibfnamefont {F.}~\bibnamefont {Dulat}},\ and\ \bibinfo {author} {\bibfnamefont {L.}~\bibnamefont {Tancredi}},\ }\bibfield  {title} {\bibinfo {title} {{Elliptic polylogarithms and iterated integrals on elliptic curves. Part I: general formalism}},\ }\href {https://doi.org/10.1007/JHEP05(2018)093} {\bibfield  {journal} {\bibinfo  {journal} {JHEP}\ }\textbf {\bibinfo {volume} {05}},\ \bibinfo {pages} {093}},\ \Eprint {https://arxiv.org/abs/1712.07089} {arXiv:1712.07089 [hep-th]} \BibitemShut {NoStop}%
\bibitem [{\citenamefont {Anastasiou}\ \emph {et~al.}(2003)\citenamefont {Anastasiou}, \citenamefont {Dixon}, \citenamefont {Bern},\ and\ \citenamefont {Kosower}}]{Anastasiou:2003kj}%
  \BibitemOpen
  \bibfield  {author} {\bibinfo {author} {\bibfnamefont {C.}~\bibnamefont {Anastasiou}}, \bibinfo {author} {\bibfnamefont {L.~J.}\ \bibnamefont {Dixon}}, \bibinfo {author} {\bibfnamefont {Z.}~\bibnamefont {Bern}},\ and\ \bibinfo {author} {\bibfnamefont {D.~A.}\ \bibnamefont {Kosower}},\ }\bibfield  {title} {\bibinfo {title} {{Planar amplitudes in maximally supersymmetric Yang-Mills theory}},\ }\href {https://doi.org/10.1103/PhysRevLett.91.251602} {\bibfield  {journal} {\bibinfo  {journal} {Phys. Rev. Lett.}\ }\textbf {\bibinfo {volume} {91}},\ \bibinfo {pages} {251602} (\bibinfo {year} {2003})},\ \Eprint {https://arxiv.org/abs/hep-th/0309040} {arXiv:hep-th/0309040 [hep-th]} \BibitemShut {NoStop}%
%%CITATION = HEP-TH/0309040;%%
\bibitem [{\citenamefont {Bern}\ \emph {et~al.}(2005)\citenamefont {Bern}, \citenamefont {Dixon},\ and\ \citenamefont {Smirnov}}]{Bern:2005iz}%
  \BibitemOpen
  \bibfield  {author} {\bibinfo {author} {\bibfnamefont {Z.}~\bibnamefont {Bern}}, \bibinfo {author} {\bibfnamefont {L.~J.}\ \bibnamefont {Dixon}},\ and\ \bibinfo {author} {\bibfnamefont {V.~A.}\ \bibnamefont {Smirnov}},\ }\bibfield  {title} {\bibinfo {title} {{Iteration of planar amplitudes in maximally supersymmetric Yang-Mills theory at three loops and beyond}},\ }\href {https://doi.org/10.1103/PhysRevD.72.085001} {\bibfield  {journal} {\bibinfo  {journal} {Phys. Rev.}\ }\textbf {\bibinfo {volume} {D72}},\ \bibinfo {pages} {085001} (\bibinfo {year} {2005})},\ \Eprint {https://arxiv.org/abs/hep-th/0505205} {arXiv:hep-th/0505205 [hep-th]} \BibitemShut {NoStop}%
%%CITATION = HEP-TH/0505205;%%
\bibitem [{\citenamefont {Caron-Huot}\ and\ \citenamefont {Larsen}(2012)}]{CaronHuot:2012ab}%
  \BibitemOpen
  \bibfield  {author} {\bibinfo {author} {\bibfnamefont {S.}~\bibnamefont {Caron-Huot}}\ and\ \bibinfo {author} {\bibfnamefont {K.~J.}\ \bibnamefont {Larsen}},\ }\bibfield  {title} {\bibinfo {title} {{Uniqueness of two-loop master contours}},\ }\href {https://doi.org/10.1007/JHEP10(2012)026} {\bibfield  {journal} {\bibinfo  {journal} {JHEP}\ }\textbf {\bibinfo {volume} {10}},\ \bibinfo {pages} {026}},\ \Eprint {https://arxiv.org/abs/1205.0801} {arXiv:1205.0801 [hep-ph]} \BibitemShut {NoStop}%
\bibitem [{\citenamefont {Paulos}\ \emph {et~al.}(2012)\citenamefont {Paulos}, \citenamefont {Spradlin},\ and\ \citenamefont {Volovich}}]{Paulos:2012nu}%
  \BibitemOpen
  \bibfield  {author} {\bibinfo {author} {\bibfnamefont {M.~F.}\ \bibnamefont {Paulos}}, \bibinfo {author} {\bibfnamefont {M.}~\bibnamefont {Spradlin}},\ and\ \bibinfo {author} {\bibfnamefont {A.}~\bibnamefont {Volovich}},\ }\bibfield  {title} {\bibinfo {title} {{Mellin Amplitudes for Dual Conformal Integrals}},\ }\href {https://doi.org/10.1007/JHEP08(2012)072} {\bibfield  {journal} {\bibinfo  {journal} {JHEP}\ }\textbf {\bibinfo {volume} {08}},\ \bibinfo {pages} {072}},\ \Eprint {https://arxiv.org/abs/1203.6362} {arXiv:1203.6362 [hep-th]} \BibitemShut {NoStop}%
\bibitem [{\citenamefont {Nandan}\ \emph {et~al.}(2013)\citenamefont {Nandan}, \citenamefont {Paulos}, \citenamefont {Spradlin},\ and\ \citenamefont {Volovich}}]{Nandan:2013ip}%
  \BibitemOpen
  \bibfield  {author} {\bibinfo {author} {\bibfnamefont {D.}~\bibnamefont {Nandan}}, \bibinfo {author} {\bibfnamefont {M.~F.}\ \bibnamefont {Paulos}}, \bibinfo {author} {\bibfnamefont {M.}~\bibnamefont {Spradlin}},\ and\ \bibinfo {author} {\bibfnamefont {A.}~\bibnamefont {Volovich}},\ }\bibfield  {title} {\bibinfo {title} {{Star Integrals, Convolutions and Simplices}},\ }\href {https://doi.org/10.1007/JHEP05(2013)105} {\bibfield  {journal} {\bibinfo  {journal} {JHEP}\ }\textbf {\bibinfo {volume} {05}},\ \bibinfo {pages} {105}},\ \Eprint {https://arxiv.org/abs/1301.2500} {arXiv:1301.2500 [hep-th]} \BibitemShut {NoStop}%
\bibitem [{\citenamefont {Bourjaily}\ \emph {et~al.}(2018)\citenamefont {Bourjaily}, \citenamefont {McLeod}, \citenamefont {Spradlin}, \citenamefont {von Hippel},\ and\ \citenamefont {Wilhelm}}]{Bourjaily:2017bsb}%
  \BibitemOpen
  \bibfield  {author} {\bibinfo {author} {\bibfnamefont {J.~L.}\ \bibnamefont {Bourjaily}}, \bibinfo {author} {\bibfnamefont {A.~J.}\ \bibnamefont {McLeod}}, \bibinfo {author} {\bibfnamefont {M.}~\bibnamefont {Spradlin}}, \bibinfo {author} {\bibfnamefont {M.}~\bibnamefont {von Hippel}},\ and\ \bibinfo {author} {\bibfnamefont {M.}~\bibnamefont {Wilhelm}},\ }\bibfield  {title} {\bibinfo {title} {{Elliptic Double-Box Integrals: Massless Scattering Amplitudes beyond Polylogarithms}},\ }\href {https://doi.org/10.1103/PhysRevLett.120.121603} {\bibfield  {journal} {\bibinfo  {journal} {Phys. Rev. Lett.}\ }\textbf {\bibinfo {volume} {120}},\ \bibinfo {pages} {121603} (\bibinfo {year} {2018})},\ \Eprint {https://arxiv.org/abs/1712.02785} {arXiv:1712.02785 [hep-th]} \BibitemShut {NoStop}%
%%CITATION = ARXIV:1712.02785;%%
\bibitem [{\citenamefont {Kristensson}\ \emph {et~al.}(2021)\citenamefont {Kristensson}, \citenamefont {Wilhelm},\ and\ \citenamefont {Zhang}}]{Kristensson:2021ani}%
  \BibitemOpen
  \bibfield  {author} {\bibinfo {author} {\bibfnamefont {A.}~\bibnamefont {Kristensson}}, \bibinfo {author} {\bibfnamefont {M.}~\bibnamefont {Wilhelm}},\ and\ \bibinfo {author} {\bibfnamefont {C.}~\bibnamefont {Zhang}},\ }\bibfield  {title} {\bibinfo {title} {{Elliptic Double Box and Symbology Beyond Polylogarithms}},\ }\href {https://doi.org/10.1103/PhysRevLett.127.251603} {\bibfield  {journal} {\bibinfo  {journal} {Phys. Rev. Lett.}\ }\textbf {\bibinfo {volume} {127}},\ \bibinfo {pages} {251603} (\bibinfo {year} {2021})},\ \Eprint {https://arxiv.org/abs/2106.14902} {arXiv:2106.14902 [hep-th]} \BibitemShut {NoStop}%
\bibitem [{\citenamefont {Goncharov}\ \emph {et~al.}(2010)\citenamefont {Goncharov}, \citenamefont {Spradlin}, \citenamefont {Vergu},\ and\ \citenamefont {Volovich}}]{Goncharov:2010jf}%
  \BibitemOpen
  \bibfield  {author} {\bibinfo {author} {\bibfnamefont {A.~B.}\ \bibnamefont {Goncharov}}, \bibinfo {author} {\bibfnamefont {M.}~\bibnamefont {Spradlin}}, \bibinfo {author} {\bibfnamefont {C.}~\bibnamefont {Vergu}},\ and\ \bibinfo {author} {\bibfnamefont {A.}~\bibnamefont {Volovich}},\ }\bibfield  {title} {\bibinfo {title} {{Classical Polylogarithms for Amplitudes and Wilson Loops}},\ }\href {https://doi.org/10.1103/PhysRevLett.105.151605} {\bibfield  {journal} {\bibinfo  {journal} {Phys. Rev. Lett.}\ }\textbf {\bibinfo {volume} {105}},\ \bibinfo {pages} {151605} (\bibinfo {year} {2010})},\ \Eprint {https://arxiv.org/abs/1006.5703} {arXiv:1006.5703 [hep-th]} \BibitemShut {NoStop}%
%%CITATION = ARXIV:1006.5703;%%
\bibitem [{\citenamefont {Broedel}\ \emph {et~al.}(2018{\natexlab{c}})\citenamefont {Broedel}, \citenamefont {Duhr}, \citenamefont {Dulat}, \citenamefont {Penante},\ and\ \citenamefont {Tancredi}}]{Broedel:2018iwv}%
  \BibitemOpen
  \bibfield  {author} {\bibinfo {author} {\bibfnamefont {J.}~\bibnamefont {Broedel}}, \bibinfo {author} {\bibfnamefont {C.}~\bibnamefont {Duhr}}, \bibinfo {author} {\bibfnamefont {F.}~\bibnamefont {Dulat}}, \bibinfo {author} {\bibfnamefont {B.}~\bibnamefont {Penante}},\ and\ \bibinfo {author} {\bibfnamefont {L.}~\bibnamefont {Tancredi}},\ }\bibfield  {title} {\bibinfo {title} {{Elliptic symbol calculus: from elliptic polylogarithms to iterated integrals of Eisenstein series}},\ }\href {https://doi.org/10.1007/JHEP08(2018)014} {\bibfield  {journal} {\bibinfo  {journal} {JHEP}\ }\textbf {\bibinfo {volume} {08}},\ \bibinfo {pages} {014}},\ \Eprint {https://arxiv.org/abs/1803.10256} {arXiv:1803.10256 [hep-th]} \BibitemShut {NoStop}%
\bibitem [{\citenamefont {Wilhelm}\ and\ \citenamefont {Zhang}(2023)}]{Wilhelm:2022wow}%
  \BibitemOpen
  \bibfield  {author} {\bibinfo {author} {\bibfnamefont {M.}~\bibnamefont {Wilhelm}}\ and\ \bibinfo {author} {\bibfnamefont {C.}~\bibnamefont {Zhang}},\ }\bibfield  {title} {\bibinfo {title} {{Symbology for elliptic multiple polylogarithms and the symbol prime}},\ }\href {https://doi.org/10.1007/JHEP01(2023)089} {\bibfield  {journal} {\bibinfo  {journal} {JHEP}\ }\textbf {\bibinfo {volume} {01}},\ \bibinfo {pages} {089}},\ \Eprint {https://arxiv.org/abs/2206.08378} {arXiv:2206.08378 [hep-th]} \BibitemShut {NoStop}%
\bibitem [{\citenamefont {Bourjaily}\ and\ \citenamefont {Trnka}(2015)}]{Bourjaily:2015jna}%
  \BibitemOpen
  \bibfield  {author} {\bibinfo {author} {\bibfnamefont {J.~L.}\ \bibnamefont {Bourjaily}}\ and\ \bibinfo {author} {\bibfnamefont {J.}~\bibnamefont {Trnka}},\ }\bibfield  {title} {\bibinfo {title} {{Local Integrand Representations of All Two-Loop Amplitudes in Planar SYM}},\ }\href {https://doi.org/10.1007/JHEP08(2015)119} {\bibfield  {journal} {\bibinfo  {journal} {JHEP}\ }\textbf {\bibinfo {volume} {08}},\ \bibinfo {pages} {119}},\ \Eprint {https://arxiv.org/abs/1505.05886} {arXiv:1505.05886 [hep-th]} \BibitemShut {NoStop}%
\bibitem [{\citenamefont {Bourjaily}\ \emph {et~al.}(2017)\citenamefont {Bourjaily}, \citenamefont {Herrmann},\ and\ \citenamefont {Trnka}}]{Bourjaily:2017wjl}%
  \BibitemOpen
  \bibfield  {author} {\bibinfo {author} {\bibfnamefont {J.~L.}\ \bibnamefont {Bourjaily}}, \bibinfo {author} {\bibfnamefont {E.}~\bibnamefont {Herrmann}},\ and\ \bibinfo {author} {\bibfnamefont {J.}~\bibnamefont {Trnka}},\ }\bibfield  {title} {\bibinfo {title} {{Prescriptive Unitarity}},\ }\href {https://doi.org/10.1007/JHEP06(2017)059} {\bibfield  {journal} {\bibinfo  {journal} {JHEP}\ }\textbf {\bibinfo {volume} {06}},\ \bibinfo {pages} {059}},\ \Eprint {https://arxiv.org/abs/1704.05460} {arXiv:1704.05460 [hep-th]} \BibitemShut {NoStop}%
\bibitem [{\citenamefont {Ren}\ \emph {et~al.}(2024)\citenamefont {Ren}, \citenamefont {Spradlin}, \citenamefont {Vergu},\ and\ \citenamefont {Volovich}}]{Ren:2023tuj}%
  \BibitemOpen
  \bibfield  {author} {\bibinfo {author} {\bibfnamefont {L.}~\bibnamefont {Ren}}, \bibinfo {author} {\bibfnamefont {M.}~\bibnamefont {Spradlin}}, \bibinfo {author} {\bibfnamefont {C.}~\bibnamefont {Vergu}},\ and\ \bibinfo {author} {\bibfnamefont {A.}~\bibnamefont {Volovich}},\ }\bibfield  {title} {\bibinfo {title} {{One-loop integrals from volumes of orthoschemes}},\ }\href {https://doi.org/10.1007/JHEP05(2024)104} {\bibfield  {journal} {\bibinfo  {journal} {JHEP}\ }\textbf {\bibinfo {volume} {05}},\ \bibinfo {pages} {104}},\ \Eprint {https://arxiv.org/abs/2306.04630} {arXiv:2306.04630 [hep-th]} \BibitemShut {NoStop}%
\bibitem [{\citenamefont {Morales}\ \emph {et~al.}(2023)\citenamefont {Morales}, \citenamefont {Spiering}, \citenamefont {Wilhelm}, \citenamefont {Yang},\ and\ \citenamefont {Zhang}}]{Morales:2022csr}%
  \BibitemOpen
  \bibfield  {author} {\bibinfo {author} {\bibfnamefont {R.}~\bibnamefont {Morales}}, \bibinfo {author} {\bibfnamefont {A.}~\bibnamefont {Spiering}}, \bibinfo {author} {\bibfnamefont {M.}~\bibnamefont {Wilhelm}}, \bibinfo {author} {\bibfnamefont {Q.}~\bibnamefont {Yang}},\ and\ \bibinfo {author} {\bibfnamefont {C.}~\bibnamefont {Zhang}},\ }\bibfield  {title} {\bibinfo {title} {{Bootstrapping Elliptic Feynman Integrals Using Schubert Analysis}},\ }\href {https://doi.org/10.1103/PhysRevLett.131.041601} {\bibfield  {journal} {\bibinfo  {journal} {Phys. Rev. Lett.}\ }\textbf {\bibinfo {volume} {131}},\ \bibinfo {pages} {041601} (\bibinfo {year} {2023})},\ \Eprint {https://arxiv.org/abs/2212.09762} {arXiv:2212.09762 [hep-th]} \BibitemShut {NoStop}%
\bibitem [{\citenamefont {Drummond}\ \emph {et~al.}(2007)\citenamefont {Drummond}, \citenamefont {Henn}, \citenamefont {Smirnov},\ and\ \citenamefont {Sokatchev}}]{Drummond:2006rz}%
  \BibitemOpen
  \bibfield  {author} {\bibinfo {author} {\bibfnamefont {J.~M.}\ \bibnamefont {Drummond}}, \bibinfo {author} {\bibfnamefont {J.}~\bibnamefont {Henn}}, \bibinfo {author} {\bibfnamefont {V.~A.}\ \bibnamefont {Smirnov}},\ and\ \bibinfo {author} {\bibfnamefont {E.}~\bibnamefont {Sokatchev}},\ }\bibfield  {title} {\bibinfo {title} {{Magic identities for conformal four-point integrals}},\ }\href {https://doi.org/10.1088/1126-6708/2007/01/064} {\bibfield  {journal} {\bibinfo  {journal} {JHEP}\ }\textbf {\bibinfo {volume} {01}},\ \bibinfo {pages} {064}},\ \Eprint {https://arxiv.org/abs/hep-th/0607160} {arXiv:hep-th/0607160 [hep-th]} \BibitemShut {NoStop}%
%%CITATION = HEP-TH/0607160;%%
\bibitem [{\citenamefont {Mack}\ and\ \citenamefont {Salam}(1969)}]{Mack:1969rr}%
  \BibitemOpen
  \bibfield  {author} {\bibinfo {author} {\bibfnamefont {G.}~\bibnamefont {Mack}}\ and\ \bibinfo {author} {\bibfnamefont {A.}~\bibnamefont {Salam}},\ }\bibfield  {title} {\bibinfo {title} {{Finite component field representations of the conformal group}},\ }\href {https://doi.org/10.1016/0003-4916(69)90278-4} {\bibfield  {journal} {\bibinfo  {journal} {Annals Phys.}\ }\textbf {\bibinfo {volume} {53}},\ \bibinfo {pages} {174} (\bibinfo {year} {1969})}\BibitemShut {NoStop}%
\bibitem [{\citenamefont {Simmons-Duffin}(2014)}]{Simmons-Duffin:2012juh}%
  \BibitemOpen
  \bibfield  {author} {\bibinfo {author} {\bibfnamefont {D.}~\bibnamefont {Simmons-Duffin}},\ }\bibfield  {title} {\bibinfo {title} {{Projectors, Shadows, and Conformal Blocks}},\ }\href {https://doi.org/10.1007/JHEP04(2014)146} {\bibfield  {journal} {\bibinfo  {journal} {JHEP}\ }\textbf {\bibinfo {volume} {04}},\ \bibinfo {pages} {146}},\ \Eprint {https://arxiv.org/abs/1204.3894} {arXiv:1204.3894 [hep-th]} \BibitemShut {NoStop}%
\bibitem [{\citenamefont {Arkani-Hamed}\ and\ \citenamefont {Yuan}(2017)}]{Arkani-Hamed:2017ahv}%
  \BibitemOpen
  \bibfield  {author} {\bibinfo {author} {\bibfnamefont {N.}~\bibnamefont {Arkani-Hamed}}\ and\ \bibinfo {author} {\bibfnamefont {E.~Y.}\ \bibnamefont {Yuan}},\ }\bibfield  {title} {\bibinfo {title} {{One-Loop Integrals from Spherical Projections of Planes and Quadrics}},\ }\href@noop {} {\  (\bibinfo {year} {2017})},\ \Eprint {https://arxiv.org/abs/1712.09991} {arXiv:1712.09991 [hep-th]} \BibitemShut {NoStop}%
\bibitem [{\citenamefont {Bourjaily}\ \emph {et~al.}(2019)\citenamefont {Bourjaily}, \citenamefont {Dulat},\ and\ \citenamefont {Panzer}}]{Bourjaily:2019jrk}%
  \BibitemOpen
  \bibfield  {author} {\bibinfo {author} {\bibfnamefont {J.~L.}\ \bibnamefont {Bourjaily}}, \bibinfo {author} {\bibfnamefont {F.}~\bibnamefont {Dulat}},\ and\ \bibinfo {author} {\bibfnamefont {E.}~\bibnamefont {Panzer}},\ }\bibfield  {title} {\bibinfo {title} {{Manifestly Dual-Conformal Loop Integration}},\ }\href {https://doi.org/10.1016/j.nuclphysb.2019.03.022} {\bibfield  {journal} {\bibinfo  {journal} {Nucl. Phys.}\ }\textbf {\bibinfo {volume} {B942}},\ \bibinfo {pages} {251} (\bibinfo {year} {2019})},\ \Eprint {https://arxiv.org/abs/1901.02887} {arXiv:1901.02887 [hep-th]} \BibitemShut {NoStop}%
%%CITATION = ARXIV:1901.02887;%%
\bibitem [{\citenamefont {Gaiotto}\ \emph {et~al.}(2011)\citenamefont {Gaiotto}, \citenamefont {Maldacena}, \citenamefont {Sever},\ and\ \citenamefont {Vieira}}]{Gaiotto:2011dt}%
  \BibitemOpen
  \bibfield  {author} {\bibinfo {author} {\bibfnamefont {D.}~\bibnamefont {Gaiotto}}, \bibinfo {author} {\bibfnamefont {J.}~\bibnamefont {Maldacena}}, \bibinfo {author} {\bibfnamefont {A.}~\bibnamefont {Sever}},\ and\ \bibinfo {author} {\bibfnamefont {P.}~\bibnamefont {Vieira}},\ }\bibfield  {title} {\bibinfo {title} {{Pulling the straps of polygons}},\ }\href {https://doi.org/10.1007/JHEP12(2011)011} {\bibfield  {journal} {\bibinfo  {journal} {JHEP}\ }\textbf {\bibinfo {volume} {12}},\ \bibinfo {pages} {011}},\ \Eprint {https://arxiv.org/abs/1102.0062} {arXiv:1102.0062 [hep-th]} \BibitemShut {NoStop}%
%%CITATION = ARXIV:1102.0062;%%
\bibitem [{\citenamefont {Davydychev}\ and\ \citenamefont {Delbourgo}(1998)}]{Davydychev:1997wa}%
  \BibitemOpen
  \bibfield  {author} {\bibinfo {author} {\bibfnamefont {A.~I.}\ \bibnamefont {Davydychev}}\ and\ \bibinfo {author} {\bibfnamefont {R.}~\bibnamefont {Delbourgo}},\ }\bibfield  {title} {\bibinfo {title} {{A Geometrical angle on Feynman integrals}},\ }\href {https://doi.org/10.1063/1.532513} {\bibfield  {journal} {\bibinfo  {journal} {J. Math. Phys.}\ }\textbf {\bibinfo {volume} {39}},\ \bibinfo {pages} {4299} (\bibinfo {year} {1998})},\ \Eprint {https://arxiv.org/abs/hep-th/9709216} {arXiv:hep-th/9709216} \BibitemShut {NoStop}%
\bibitem [{\citenamefont {Zagier}(2007)}]{Zagier2007}%
  \BibitemOpen
  \bibfield  {author} {\bibinfo {author} {\bibfnamefont {D.}~\bibnamefont {Zagier}},\ }\bibinfo {title} {The dilogarithm function},\ in\ \href {https://doi.org/10.1007/978-3-540-30308-4_1} {\emph {\bibinfo {booktitle} {Frontiers in Number Theory, Physics, and Geometry II: On Conformal Field Theories, Discrete Groups and Renormalization}}},\ \bibinfo {editor} {edited by\ \bibinfo {editor} {\bibfnamefont {P.}~\bibnamefont {Cartier}}, \bibinfo {editor} {\bibfnamefont {P.}~\bibnamefont {Moussa}}, \bibinfo {editor} {\bibfnamefont {B.}~\bibnamefont {Julia}},\ and\ \bibinfo {editor} {\bibfnamefont {P.}~\bibnamefont {Vanhove}}}\ (\bibinfo  {publisher} {Springer Berlin Heidelberg},\ \bibinfo {address} {Berlin, Heidelberg},\ \bibinfo {year} {2007})\ pp.\ \bibinfo {pages} {3--65}\BibitemShut {NoStop}%
\bibitem [{\citenamefont {Goncharov}(2009)}]{Goncharov:2009lql}%
  \BibitemOpen
  \bibfield  {author} {\bibinfo {author} {\bibfnamefont {A.~B.}\ \bibnamefont {Goncharov}},\ }\bibfield  {title} {\bibinfo {title} {{A simple construction of Grassmannian polylogarithms}},\ }\href@noop {} {\  (\bibinfo {year} {2009})},\ \Eprint {https://arxiv.org/abs/0908.2238} {arXiv:0908.2238 [math.AG]} \BibitemShut {NoStop}%
\bibitem [{\citenamefont {Duhr}\ \emph {et~al.}(2012)\citenamefont {Duhr}, \citenamefont {Gangl},\ and\ \citenamefont {Rhodes}}]{Duhr:2011zq}%
  \BibitemOpen
  \bibfield  {author} {\bibinfo {author} {\bibfnamefont {C.}~\bibnamefont {Duhr}}, \bibinfo {author} {\bibfnamefont {H.}~\bibnamefont {Gangl}},\ and\ \bibinfo {author} {\bibfnamefont {J.~R.}\ \bibnamefont {Rhodes}},\ }\bibfield  {title} {\bibinfo {title} {{From polygons and symbols to polylogarithmic functions}},\ }\href {https://doi.org/10.1007/JHEP10(2012)075} {\bibfield  {journal} {\bibinfo  {journal} {JHEP}\ }\textbf {\bibinfo {volume} {10}},\ \bibinfo {pages} {075}},\ \Eprint {https://arxiv.org/abs/1110.0458} {arXiv:1110.0458 [math-ph]} \BibitemShut {NoStop}%
%%CITATION = ARXIV:1110.0458;%%
\bibitem [{\citenamefont {Duhr}(2012)}]{Duhr:2012fh}%
  \BibitemOpen
  \bibfield  {author} {\bibinfo {author} {\bibfnamefont {C.}~\bibnamefont {Duhr}},\ }\bibfield  {title} {\bibinfo {title} {{Hopf algebras, coproducts and symbols: an application to Higgs boson amplitudes}},\ }\href {https://doi.org/10.1007/JHEP08(2012)043} {\bibfield  {journal} {\bibinfo  {journal} {JHEP}\ }\textbf {\bibinfo {volume} {08}},\ \bibinfo {pages} {043}},\ \Eprint {https://arxiv.org/abs/1203.0454} {arXiv:1203.0454 [hep-ph]} \BibitemShut {NoStop}%
%%CITATION = ARXIV:1203.0454;%%
\bibitem [{\citenamefont {He}\ and\ \citenamefont {Tang}(2023)}]{He:2023qld}%
  \BibitemOpen
  \bibfield  {author} {\bibinfo {author} {\bibfnamefont {S.}~\bibnamefont {He}}\ and\ \bibinfo {author} {\bibfnamefont {Y.}~\bibnamefont {Tang}},\ }\bibfield  {title} {\bibinfo {title} {{Algorithm for symbol integrations for loop integrals}},\ }\href {https://doi.org/10.1103/PhysRevD.108.L041702} {\bibfield  {journal} {\bibinfo  {journal} {Phys. Rev. D}\ }\textbf {\bibinfo {volume} {108}},\ \bibinfo {pages} {L041702} (\bibinfo {year} {2023})},\ \Eprint {https://arxiv.org/abs/2304.01776} {arXiv:2304.01776 [hep-th]} \BibitemShut {NoStop}%
\bibitem [{\citenamefont {Goncharov}(2005)}]{Gonch2}%
  \BibitemOpen
  \bibfield  {author} {\bibinfo {author} {\bibfnamefont {A.~B.}\ \bibnamefont {Goncharov}},\ }\bibfield  {title} {\bibinfo {title} {Galois symmetries of fundamental groupoids and noncommutative geometry},\ }\href {https://doi.org/10.1215/S0012-7094-04-12822-2} {\bibfield  {journal} {\bibinfo  {journal} {Duke Math. J.}\ }\textbf {\bibinfo {volume} {128}},\ \bibinfo {pages} {209} (\bibinfo {year} {2005})},\ \Eprint {https://arxiv.org/abs/math/0208144} {arXiv:math/0208144 [math.AG]} \BibitemShut {NoStop}%
\bibitem [{\citenamefont {Goncharov}(2001)}]{Goncharov:2001iea}%
  \BibitemOpen
  \bibfield  {author} {\bibinfo {author} {\bibfnamefont {A.~B.}\ \bibnamefont {Goncharov}},\ }\bibfield  {title} {\bibinfo {title} {{Multiple polylogarithms and mixed Tate motives}},\ }\href@noop {} {\  (\bibinfo {year} {2001})},\ \Eprint {https://arxiv.org/abs/math/0103059} {arXiv:math/0103059} \BibitemShut {NoStop}%
\bibitem [{\citenamefont {Spradlin}\ and\ \citenamefont {Volovich}(2011)}]{Spradlin:2011wp}%
  \BibitemOpen
  \bibfield  {author} {\bibinfo {author} {\bibfnamefont {M.}~\bibnamefont {Spradlin}}\ and\ \bibinfo {author} {\bibfnamefont {A.}~\bibnamefont {Volovich}},\ }\bibfield  {title} {\bibinfo {title} {{Symbols of One-Loop Integrals From Mixed Tate Motives}},\ }\href {https://doi.org/10.1007/JHEP11(2011)084} {\bibfield  {journal} {\bibinfo  {journal} {JHEP}\ }\textbf {\bibinfo {volume} {11}},\ \bibinfo {pages} {084}},\ \Eprint {https://arxiv.org/abs/1105.2024} {arXiv:1105.2024 [hep-th]} \BibitemShut {NoStop}%
\bibitem [{\citenamefont {Caron-Huot}(2011)}]{Caron-Huot:2011zgw}%
  \BibitemOpen
  \bibfield  {author} {\bibinfo {author} {\bibfnamefont {S.}~\bibnamefont {Caron-Huot}},\ }\bibfield  {title} {\bibinfo {title} {{Superconformal symmetry and two-loop amplitudes in planar $\mathcal{N}=4$ super Yang-Mills}},\ }\href {https://doi.org/10.1007/JHEP12(2011)066} {\bibfield  {journal} {\bibinfo  {journal} {JHEP}\ }\textbf {\bibinfo {volume} {12}},\ \bibinfo {pages} {066}},\ \Eprint {https://arxiv.org/abs/1105.5606} {arXiv:1105.5606 [hep-th]} \BibitemShut {NoStop}%
\bibitem [{\citenamefont {He}\ \emph {et~al.}(2022{\natexlab{a}})\citenamefont {He}, \citenamefont {Li},\ and\ \citenamefont {Yang}}]{He:2021mme}%
  \BibitemOpen
  \bibfield  {author} {\bibinfo {author} {\bibfnamefont {S.}~\bibnamefont {He}}, \bibinfo {author} {\bibfnamefont {Z.}~\bibnamefont {Li}},\ and\ \bibinfo {author} {\bibfnamefont {Q.}~\bibnamefont {Yang}},\ }\bibfield  {title} {\bibinfo {title} {{Comments on all-loop constraints for scattering amplitudes and Feynman integrals}},\ }\href {https://doi.org/10.1007/JHEP01(2022)073} {\bibfield  {journal} {\bibinfo  {journal} {JHEP}\ }\textbf {\bibinfo {volume} {01}},\ \bibinfo {pages} {073}},\ \bibinfo {note} {[Erratum: JHEP 05, 076 (2022)]},\ \Eprint {https://arxiv.org/abs/2108.07959} {arXiv:2108.07959 [hep-th]} \BibitemShut {NoStop}%
\bibitem [{\citenamefont {Broedel}\ \emph {et~al.}(2019)\citenamefont {Broedel}, \citenamefont {Duhr}, \citenamefont {Dulat}, \citenamefont {Penante},\ and\ \citenamefont {Tancredi}}]{Broedel:2018qkq}%
  \BibitemOpen
  \bibfield  {author} {\bibinfo {author} {\bibfnamefont {J.}~\bibnamefont {Broedel}}, \bibinfo {author} {\bibfnamefont {C.}~\bibnamefont {Duhr}}, \bibinfo {author} {\bibfnamefont {F.}~\bibnamefont {Dulat}}, \bibinfo {author} {\bibfnamefont {B.}~\bibnamefont {Penante}},\ and\ \bibinfo {author} {\bibfnamefont {L.}~\bibnamefont {Tancredi}},\ }\bibfield  {title} {\bibinfo {title} {{Elliptic Feynman integrals and pure functions}},\ }\href {https://doi.org/10.1007/JHEP01(2019)023} {\bibfield  {journal} {\bibinfo  {journal} {JHEP}\ }\textbf {\bibinfo {volume} {01}},\ \bibinfo {pages} {023}},\ \Eprint {https://arxiv.org/abs/1809.10698} {arXiv:1809.10698 [hep-th]} \BibitemShut {NoStop}%
%%CITATION = ARXIV:1809.10698;%%
\bibitem [{\citenamefont {Chen}(1977)}]{Chen:1977oja}%
  \BibitemOpen
  \bibfield  {author} {\bibinfo {author} {\bibfnamefont {K.-T.}\ \bibnamefont {Chen}},\ }\bibfield  {title} {\bibinfo {title} {{Iterated path integrals}},\ }\href {https://doi.org/10.1090/S0002-9904-1977-14320-6} {\bibfield  {journal} {\bibinfo  {journal} {Bull. Am. Math. Soc.}\ }\textbf {\bibinfo {volume} {83}},\ \bibinfo {pages} {831} (\bibinfo {year} {1977})}\BibitemShut {NoStop}%
%%CITATION = BAMOA,83,831;%%
\bibitem [{\citenamefont {Brown}(2009)}]{Brown:2009qja}%
  \BibitemOpen
  \bibfield  {author} {\bibinfo {author} {\bibfnamefont {F.~C.~S.}\ \bibnamefont {Brown}},\ }\bibfield  {title} {\bibinfo {title} {{Multiple zeta values and periods of moduli spaces $\overline{\mathfrak{M}}_{0,n}(\mathbb{R})$}},\ }\href@noop {} {\bibfield  {journal} {\bibinfo  {journal} {Annales Sci. Ecole Norm. Sup.}\ }\textbf {\bibinfo {volume} {42}},\ \bibinfo {pages} {371} (\bibinfo {year} {2009})},\ \Eprint {https://arxiv.org/abs/math/0606419} {arXiv:math/0606419} \BibitemShut {NoStop}%
\bibitem [{\citenamefont {Henn}\ \emph {et~al.}(2024)\citenamefont {Henn}, \citenamefont {Matija\v{s}i\'c}, \citenamefont {Miczajka}, \citenamefont {Peraro}, \citenamefont {Xu},\ and\ \citenamefont {Zhang}}]{Henn:2024ngj}%
  \BibitemOpen
  \bibfield  {author} {\bibinfo {author} {\bibfnamefont {J.~M.}\ \bibnamefont {Henn}}, \bibinfo {author} {\bibfnamefont {A.}~\bibnamefont {Matija\v{s}i\'c}}, \bibinfo {author} {\bibfnamefont {J.}~\bibnamefont {Miczajka}}, \bibinfo {author} {\bibfnamefont {T.}~\bibnamefont {Peraro}}, \bibinfo {author} {\bibfnamefont {Y.}~\bibnamefont {Xu}},\ and\ \bibinfo {author} {\bibfnamefont {Y.}~\bibnamefont {Zhang}},\ }\bibfield  {title} {\bibinfo {title} {{A computation of two-loop six-point Feynman integrals in dimensional regularization}},\ }\href@noop {} {\  (\bibinfo {year} {2024})},\ \Eprint {https://arxiv.org/abs/2403.19742} {arXiv:2403.19742 [hep-ph]} \BibitemShut {NoStop}%
\bibitem [{\citenamefont {Bourjaily}\ \emph {et~al.}(2015)\citenamefont {Bourjaily}, \citenamefont {Caron-Huot},\ and\ \citenamefont {Trnka}}]{Bourjaily:2013mma}%
  \BibitemOpen
  \bibfield  {author} {\bibinfo {author} {\bibfnamefont {J.~L.}\ \bibnamefont {Bourjaily}}, \bibinfo {author} {\bibfnamefont {S.}~\bibnamefont {Caron-Huot}},\ and\ \bibinfo {author} {\bibfnamefont {J.}~\bibnamefont {Trnka}},\ }\bibfield  {title} {\bibinfo {title} {{Dual-Conformal Regularization of Infrared Loop Divergences and the Chiral Box Expansion}},\ }\href {https://doi.org/10.1007/JHEP01(2015)001} {\bibfield  {journal} {\bibinfo  {journal} {JHEP}\ }\textbf {\bibinfo {volume} {01}},\ \bibinfo {pages} {001}},\ \Eprint {https://arxiv.org/abs/1303.4734} {arXiv:1303.4734 [hep-th]} \BibitemShut {NoStop}%
%%CITATION = ARXIV:1303.4734;%%
\bibitem [{\citenamefont {He}\ \emph {et~al.}(2022{\natexlab{b}})\citenamefont {He}, \citenamefont {Li},\ and\ \citenamefont {Zhang}}]{He:2022ujv}%
  \BibitemOpen
  \bibfield  {author} {\bibinfo {author} {\bibfnamefont {S.}~\bibnamefont {He}}, \bibinfo {author} {\bibfnamefont {Z.}~\bibnamefont {Li}},\ and\ \bibinfo {author} {\bibfnamefont {C.}~\bibnamefont {Zhang}},\ }\bibfield  {title} {\bibinfo {title} {{A nice two-loop next-to-next-to-MHV amplitude in $ \mathcal{N} $ = 4 super-Yang-Mills}},\ }\href {https://doi.org/10.1007/JHEP12(2022)158} {\bibfield  {journal} {\bibinfo  {journal} {JHEP}\ }\textbf {\bibinfo {volume} {12}},\ \bibinfo {pages} {158}},\ \Eprint {https://arxiv.org/abs/2209.10856} {arXiv:2209.10856 [hep-th]} \BibitemShut {NoStop}%
\end{thebibliography}%

\end{document}